\documentclass[12pt,preprint]{aastex}
\usepackage{lscape}

\newcommand{\bdv}[1]{\mbox{\boldmath$#1$}}

\def\rel{{\rm rel}}
\def\e{{\rm E}}

\def\kms{{\rm km}\,{\rm s}^{-1}}

\def\rel{{\rm rel}}

\def\geo{{\rm geo}}

\def\bv{{\bf v}}

\def\e{{\rm E}}
\def\bpi{{\bdv{\pi}}}

\def\bmu{{\bdv{\mu}}}

\begin{document}
\title{Sub-Saturn Planet MOA-2008-BLG-310Lb: Likely To Be In The Galactic Bulge\altaffilmark{1}}

\author{
Julia~Janczak\altaffilmark{2,3},
A.~Fukui\altaffilmark{4,5},
Subo~Dong\altaffilmark{2,3},
B.~Monard\altaffilmark{6},
Szymon~Koz\l owski\altaffilmark{2,3},
A.~Gould\altaffilmark{2,3},
J.P.~Beaulieu\altaffilmark{7,8,9,10},
Daniel~Kubas\altaffilmark{7,8,9,11},
J.B.~Marquette\altaffilmark{7,8,9},
T.~Sumi\altaffilmark{4,5},
I.A.~Bond\altaffilmark{4,12},
D.P.~Bennett\altaffilmark{4,7,13}\\
and\\
F.~Abe\altaffilmark{5},
K.~Furusawa\altaffilmark{5},
J.B.~Hearnshaw\altaffilmark{14},
S.~Hosaka\altaffilmark{5},
Y.~Itow\altaffilmark{5},
K.~Kamiya\altaffilmark{5},
A.V.~Korpela\altaffilmark{15},
P.M.~Kilmartin\altaffilmark{16},
W.~Lin\altaffilmark{12},
C.H.~Ling\altaffilmark{12},
S.~Makita\altaffilmark{5},
K.~Masuda\altaffilmark{5},
Y.~Matsubara\altaffilmark{5},
N.~Miyake\altaffilmark{5},
Y.~Muraki\altaffilmark{17},
M.~Nagaya\altaffilmark{5},
T.~Nagayama\altaffilmark{18},
K.~Nishimoto\altaffilmark{5},
K.~Ohnishi\altaffilmark{19},
Y.C.~Perrott\altaffilmark{20},
N.J.~Rattenbury\altaffilmark{20},
T.~Sako\altaffilmark{5},
To.~Saito\altaffilmark{21},
L.~Skuljan\altaffilmark{12},
D.J.~Sullivan\altaffilmark{15},
W.L.~Sweatman\altaffilmark{12},
P.J.~Tristram\altaffilmark{16},
P.C.M.~Yock\altaffilmark{20}\\
(The MOA Collaboration\altaffilmark{4})\\
Jin~H.~An\altaffilmark{22,23},
G.W.~Christie\altaffilmark{24},
Sun-Ju~Chung\altaffilmark{25},
D.L.~DePoy\altaffilmark{26},
B.S.~Gaudi\altaffilmark{3},
C.~Han\altaffilmark{27},
C.-U.~Lee\altaffilmark{25},
F.~Mallia\altaffilmark{28},
T.~Natusch\altaffilmark{29},
B.-G.~Park\altaffilmark{25},
R.W.~Pogge\altaffilmark{3}\\
(The $\mu$FUN Collaboration\altaffilmark{2})\\
T.~Anguita\altaffilmark{30},
M.~Dominik\altaffilmark{31,32}, 
U.G.~J{\o}rgensen\altaffilmark{33}, 
G.~Masi\altaffilmark{34}, 
M.~Mathiasen\altaffilmark{33},
S.C.~Novati\altaffilmark{35}
(The MiNDSTEp Consortium\altaffilmark{36})\\
V.~Batista\altaffilmark{8,9},
S.~Brillant\altaffilmark{11},
A.~Cassan\altaffilmark{30},
A.~Cole\altaffilmark{37},
E.~Corrales\altaffilmark{8,9},
Ch.~Coutures\altaffilmark{8,9},
S.~Dieters\altaffilmark{8,9,37},
P.~Fouqu\'e\altaffilmark{38},
J.~Greenhill\altaffilmark{37}\\
(The PLANET Collaboration\altaffilmark{7})\\
}

\altaffiltext{1}
{Partly based on observations performed at the European Southern
Observatory Very Large Telescope on Cerro Paranal, Chile.}
\altaffiltext{2}
{Microlensing Follow Up Network ($\mu$FUN),
http://www.astronomy.ohio-state.edu/$\sim$microfun}
\altaffiltext{3}
{Department of Astronomy, Ohio State University,
140 W.\ 18th Ave., Columbus, OH 43210, USA; 
dong,gaudi,gould,simkoz,pogge,@astronomy.ohio-state.edu}
\altaffiltext{4}
{Microlensing Observations in Astrophysics (MOA) Collaboration,\\  http://www.phys.canterbury.ac.nz/moa}
\altaffiltext{5}
{Solar-Terrestrial Environment Laboratory, Nagoya University, Nagoya 464-8601, Japan}
\altaffiltext{6} 
{Bronberg Observatory, South Africa, lagmonar@nmisa.org}
\altaffiltext{7}
{Probing Lensing Anomalies Network (PLANET),
http://planet.iap.fr}
\altaffiltext{8}
{UPMC Univ Paris 06, UMR7095, Institut d'Astrophysique de Paris, F-75014, Paris, France}
\altaffiltext{9}
{CNRS, UMR7095, Institut d'Astrophysique de Paris, F-75014, Paris, France}
\altaffiltext{10}
{University College London, Gower street, London WC1E 6BT, UK}
\altaffiltext{11}
{European Southern Observatory, Casilla 19001, Vitacura 19, Santiago, Chile}
\altaffiltext{12}
{Institute of Information and Mathematical Sciences, Massey University,
Private Bag 102-904, North Shore Mail Centre, Auckland, New Zealand; i.a.bond@massey.ac.nz}
\altaffiltext{13}
{Department of Physics, University of Notre Dame, Notre Dame, IN 46556, USA; bennett@nd.edu}
\altaffiltext{14}
{Department of Physics and Astronomy, University of Canterbury, Private Bag 4800, Christchurch, New Zealand}
\altaffiltext{15}
{School of Chemical and Physical Sciences, Victoria University, Wellington, New Zealand}
\altaffiltext{16}
{Mt. John Observatory, P.O. Box 56, Lake Tekapo 8770, New Zealand}
\altaffiltext{17}
{Konan University, Kobe, Japan}
\altaffiltext{18}
{Department of Physics and Astrophysics, Faculty of Science, Nagoya University, Nagoya 464-8602, Japan}
\altaffiltext{19}
{Nagano National College of Technology, Nagano 381-8550, Japan}
\altaffiltext{20}
{Department of Physics, University of Auckland, Auckland, New Zealand}
\altaffiltext{21}
{Tokyo Metropolitan College of Aeronautics, Tokyo 116-8523, Japan}
\altaffiltext{22}
{Dark Cosmology Centre, Niels Bohr Institute, University of Copenhagen, 
Juliane Mariesvej 30, DK-2100 Copenhagen, Denmark}
\altaffiltext{23}
{Niels Bohr International Academy, Niels Bohr Institute, 
University of Copenhagen, Blegdamsvej 17, DK-2100 Copenhagen, Denmark}
\altaffiltext{24}
{Auckland Observatory, Auckland, New Zealand, gwchristie@christie.org.nz}
\altaffiltext{25}
{Korea Astronomy and Space Science Institute, Daejon 305-348, Korea; sjchung,bgpark,leecu,@kasi.re.kr}
\altaffiltext{26} 
{Department of Physics, Texas A\&M University, 4242 TAMU, College Station, TX 77843-4242, USA}
\altaffiltext{27}
{Department of Physics, Institute for Basic Science Research, Chungbuk National University, Chongju 361-763, Korea;
cheongho@astroph.chungbuk.ac.kr}
\altaffiltext{28}
{Campo Catino Austral Observatory, San Pedro de Atacama, Chile}
\altaffiltext{29} 
{Institute for Radiophysics and Space Research, AUT University, Auckland,
New Zealand, tim.natusch@aut.ac.nz}
\altaffiltext{30}
{Astronomisches Rechen-Institut, Zentrum f{\"u}r Astronomie der
Universit{\"a}t Heidelberg, M{\"o}nchhofstrasse 12-14,
69120 Heidelberg, Germany}
\altaffiltext{31}
{SUPA, University of St Andrews, School of Physics \& Astronomy,
North Haugh, St Andrews, KY16 9SS, United Kingdom}
\altaffiltext{32}
{Royal Society University Research Fellow}
\altaffiltext{33}
{Niels Bohr Institute and Centre for Stars and Planet Formation,
Juliane Mariesvej 30, 2100 Copenhagen, Denmark}
\altaffiltext{34}
{Bellatrix Observatory, Via Madonna de Loco 47, 03023 Ceccano, Italy}
\altaffiltext{35}
{Dipartimento di Fisica, Universita' di Salerno and INFN, sez. di Napoli, Italy}
\altaffiltext{36}
{Microlensing Network for the Detection of Small Terrestrial Exoplanets (MiNDSTEp), 
http://www.mindstep-science.org}
\altaffiltext{37}
{University of Tasmania, School of Math and Physics, Private bag 37, GPO Hobart, Tasmania 7001, Australia}
\altaffiltext{38}
{LATT, Universit\'e de Toulouse, CNRS, 14 av. E. Belin, 31400 Toulouse, France}

\begin{abstract}
We report the detection of sub-Saturn-mass planet MOA-2008-BLG-310Lb
and argue that it is the strongest candidate yet for a bulge
planet. Deviations from the single-lens fit are smoothed out by
finite-source effects and so are not immediately apparent from the
light curve. Nevertheless, we find that a model in which the primary
has a planetary companion is favored over the single-lens model by
$\Delta\chi^2 \sim 880$ for an additional three degrees of
freedom. Detailed analysis yields a planet/star mass ratio $q=(3.3\pm
0.3)\times 10^{-4}$ and an angular separation between the planet and
star within $10\%$ of the angular Einstein radius. The small angular
Einstein radius, $\theta_E = 0.155 \pm 0.011\,{\rm mas}$, constrains
the distance to the lens to be $D_L>6.0$\,kpc if it is a star
($M_L>0.08\,M_\odot$). This is the only microlensing exoplanet host
discovered so far that must be in the bulge if it is a star. By
analyzing VLT NACO adaptive optics images taken near the baseline of
the event, we detect additional blended light that is aligned to within 
130\,mas of the lensed source. This light is plausibly from the lens, 
but could also be due to a companion to lens or source, or possibly 
an unassociated star. 
If the blended light is indeed due to the lens, we
can estimate the mass of the lens, $M_L = 0.67\pm0.14\,M_\odot$,
planet mass $m = 74 \pm 17 M_{\oplus}$, and projected separation
between the planet and host, $1.25 \pm 0.10$\,AU, putting it
right on the ``snow line''. If not, then
the planet has lower mass, is closer to its host and is colder.  
To distinguish among these possibilities on reasonable
timescales would require obtaining {\it Hubble Space Telescope}
images almost immediately, before the source-lens relative motion of
$\mu=5\,\rm mas\,yr^{-1}$ causes them to separate substantially.

\end{abstract}

\section{Introduction
\label{sec:intro}}

Over the past 5 years, gravitational microlensing has led to the discovery of several exoplanets that would not be detectable by any other method currently available. Because it does not rely on light coming from the planet or the host star, microlensing is able to detect planets at several kpc, probing even into the center of the Galaxy. Thus microlensing has the potential to determine the demographics of planets orbiting hosts from two distinct stellar populations, bulge stars and disk stars, which is critical for understanding the Galactic distribution of planets and may allow one to constrain the time-history of planet formation in the universe. 

Standard models of the spatial and velocity distributions of stars in
the Galaxy predict that roughly $2/3$ of all microlensing events of
stars in the bulge arise from bulge lenses, i.e., stars in the bulge
\citep{kiraga94}. In light of this, one might expect that planetary
detections via microlensing would be more frequent in the bulge than
in the disk. On the contrary, of the eight microlensing planets
discovered so far
\citep{bond04,udalski05,beaulieu06,gould06,gaudi08,bennett08,dong08},
five have measured or constrained lens distances, and four of these
are clearly in the foreground disk while none are unambiguously
in the bulge. The distance
is not well constrained for the remaining three planets
(OGLE-2005-BLG-169Lb, OGLE-2005-BLG-390Lb, and OGLE-2007-BLG-400Lb),
and none have been definitively identified as bulge planets. The low
detection rate of bulge planets may arise from a selection bias that
favors the longer events that preferentially arise from disk lenses,
or it may reflect the underlying Galactic distribution of
planets. Here we present the analysis of a planetary signature in
microlensing event MOA-2008-BLG-310, the strongest candidate for a bulge
planet to date.

In Section~\ref{sec:observations}, we discuss the observations and data
reduction. In Section~\ref{sec:model}, we fit these data to single-lens
and planetary models.  In Section \ref{sec:limb darkening} we discuss
our treatment of limb darkening of the source, which is important because
it potentially affects the light curve at the times of maximum deviation
due to the planet.  We measure the Einstein radius in Section~\ref{sec:radius}
and thereby constrain a combination of the lens mass and distance.
To obtain a second, independent constraint on these two quanities,
we first search, in Section~\ref{sec:parallax}, for ``microlensing
parallax'' effects, but these prove too small to be observed.
The analysis of
images from SMARTS CTIO, VLT NACO, and IRSF, detailed in Section
\ref{sec:blend}, does however reveal excess light aligned with the
event. In Section \ref{sec:constraints} we discuss the four possible
sources of this excess light: the lens, a companion to the lens, a
companion to the source, or an ambient star.  Finally, Section
\ref{sec:discussion} describes how future observations with $HST$ or
adaptive optics (AO) may help characterize the host and its planet.

\section{Observations
\label{sec:observations}}

The Microlensing Observations in Astrophysics (MOA) collaboration detected microlensing event MOA-2008-BLG-310 [(RA,Dec)=(17:54:14.53,-34:46:40.99), (l,b)=(355.92,-4.56)] on 6 July 2008 (HJD$' \equiv$ HJD - 2450000 = 4654.458). MOA issued
a high-magnification alert two days later, about 12 hours before the event peaked. The color and magnitude of the source indicate that it is a G type star in the Galactic bulge, a result confirmed by high-resolution spectroscopy \citep{cohen09}.
 
The Microlensing Follow Up Network ($\mu$FUN) began to intensively monitor this event at HJD$'=4656.026$, 
less than 9 hours before the peak. The minimum predicted peak magnification was $A_{\rm max}>80$ but the best-fit model
at the time was consistent with formally infinite magnification, so the event was given high priority. Observations were 
taken by six observatories, 
MOA (New Zealand) 1.8\,m $I$,
$\mu$FUN Auckland (New Zealand) 0.41\,m $R$, 
$\mu$FUN Bronberg (South Africa) 0.36\,m unfiltered, 
$\mu$FUN SMARTS CTIO (Chile) 1.3\,m $I$, $V$, $H$,
MiNDSTEp La Silla (Chile) 1.54\,m $I$, and
PLANET Canopus (Tasmania) 1.0\,m $I$.
Only one observatory, $\mu$FUN Bronberg, was positioned to see the peak of the event. Nevertheless, this observatory provided very complete coverage. $\mu$FUN Bronberg took a total of 973 observations over the period $4656.21 < {\rm HJD}' < 4656.55$, recording the peak and all interesting anomalies. The high density of these observations allows us to bin the $\mu$FUN Bronberg data without compromising the time resolution. The binned data points (as seen in Fig.~\ref{fig:lightcurve}, below) occur every 2.5 minutes over the peak, whereas each planetary feature spans roughly an hour. $\mu$FUN SMARTS took a total of 49 images in the $I$-band, 275 images in the $H$-band, and 6 in the $V$-band. The $\mu$FUN SMARTS observations overlap $\mu$FUN Bronberg by about 2 hours, starting after the peak and providing additional coverage of the last planetary deviation.

For $\mu$FUN observatories, we primarily use
difference imaging analysis (DIA) \citep{wozniak00}, but also use DoPhot reductions of $\mu$FUN SMARTS
$H$-band data to investigate the light that is blended with the
source.
MOA data were reduced using the standard MOA difference imaging
analysis pipeline \citep{bond01}. PLANET Canopus data were reduced
using the pySIS2 pipeline, based on the ISIS 2 code of
\citet{alard00}. 
The error bars for all data are renormalized so that $\chi^2$
per degree of freedom for the best-fit planetary model is close to
unity.
 
Being unfiltered, the $\mu$FUN Bronberg data are subject to a differential extinction correction because the source has a different color than the mean color of the reference frame used by DIA. We measure this effect from the light curves of stable stars having the same color as the lens, and thereby remove it.  See \citet{dong09}.  

\section{Microlens Model
\label{sec:model}}

\begin{figure}
	\centering
		\includegraphics[width=1.00\textwidth]{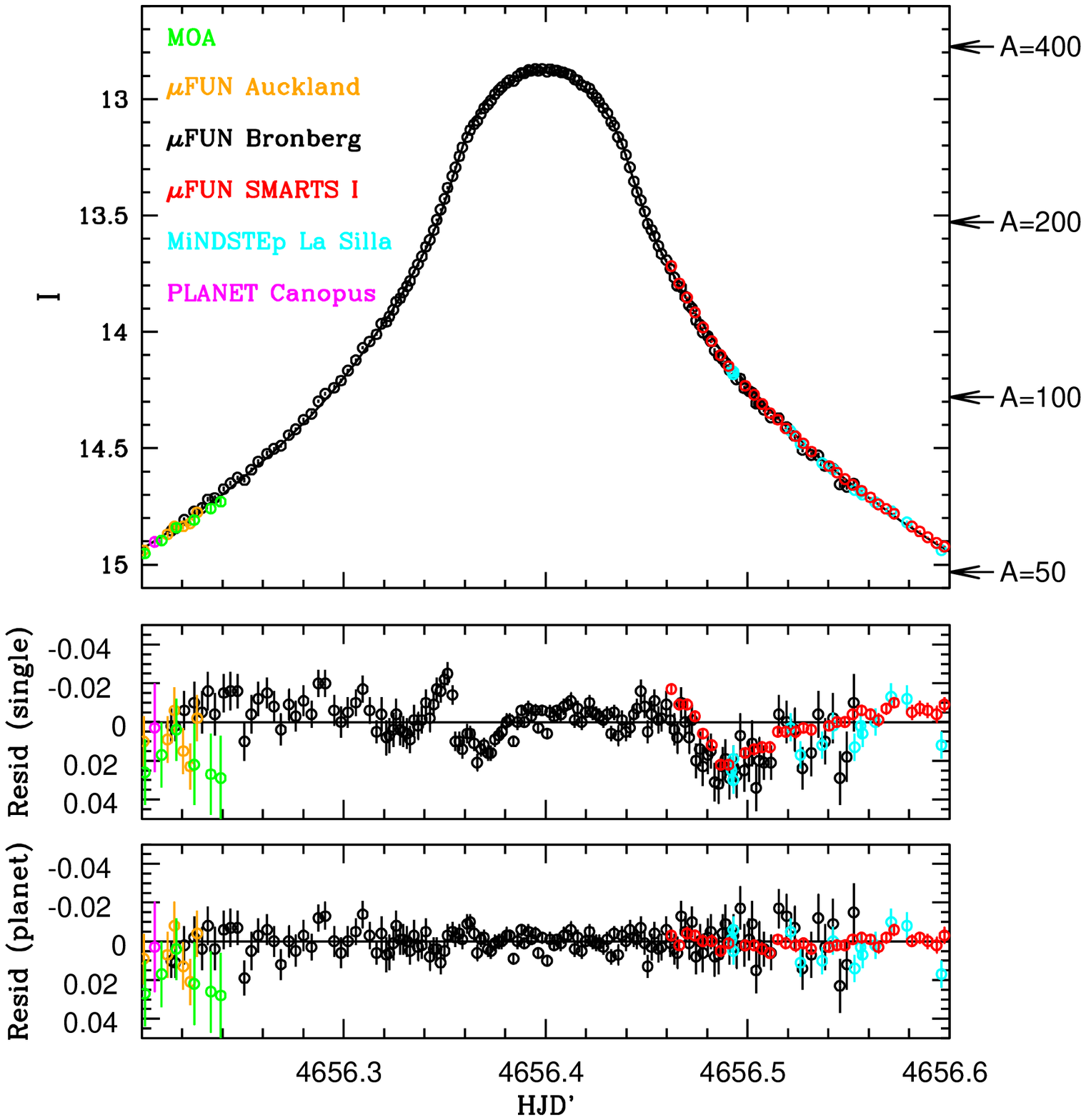}
		\caption{Top: Light curve of MOA-2008-BLG-310 showing data from MOA ({\it green}), $\mu$FUN Auckland ({\it orange}), $\mu$FUN Bronberg ({\it black}), $\mu$FUN SMARTS $I$-band ({\it red}), MiNDSTEp La Silla ({\it cyan}), and PLANET Canopus ({\it magenta}). Also shown is the best fit single-lens model. The light curve does not look anomalous at first glance. Middle: Residuals to the best-fit single-lens model. Anomalies are apparent at HJD$'= 4656.34$ and HJD$'= 4656.48$. The noticeable offset in the alignment of Bronberg and CTIO data is an effect of independently fitting $f_s$ and $f_b$ for each observatory (see Section~\ref{sec:model}). See Figure~\ref{fig:caustics} for didactic residuals. Bottom: Residuals to the best-fit planetary model (the wide solution is chosen for this plot, however the close solution is essentially indistinguishable).}
	\label{fig:lightcurve}
\end{figure}

MOA-2008-BLG-310 was initially modeled as a single lens event. The single-lens model light curve
fits the data reasonably well, showing pronounced finite-source effects in the rounding of the peak but no obvious 
anomalies. The event reached a maximum magnification $A_{\rm max}\sim 400$, making it a good candidate for planet detection
although the finite-source effects work to smooth out any planetary deviations. Figure~\ref{fig:lightcurve} shows the light curve and the residuals to the the best-fit single-lens and planetary models. The model allows us to roughly determine several parameters pertaining to the general
structure of the light curve; $t_0,u_0,t_\e$, and $\rho$. Here, $t_0$ is the time of minimum separation between the source 
and lens, $u_0$ is the minimum separation in units of the Einstein radius, $\rho$ is the radius of the source in the 
same units, and $t_{\rm E}$ is the Einstein crossing time. We find that the source crossing time, $t_*\equiv \rho t_{\rm E}$, is better constrained than $\rho$, and so we report this parameter as well. 

A close look at the residuals from the single-lens fit (middle panel of Fig.~\ref{fig:lightcurve}) reveals significant structure indicating that the underlying lens model is more complicated than a simple single lens. In particular, short timescale deviations near the peak of high-magnification events are typically caused by a planetary or binary companion. In these cases, the caustic structure is extended, as opposed to the simple point in the case of an isolated lens, leading to deviations from the single-lens form as the source crosses the caustic. The most important features, a short spike in the residuals just before the peak (HJD$'$ 4656.34), and a short dip just after (HJD$'$ 4656.48) are completely covered by unfiltered
observations from $\mu$FUN Bronberg. The second of these features is confirmed in $I$-band data from $\mu$FUN SMARTS. 
SMARTS $H$-band observations qualitatively show the same deviation despite suffering from larger scatter. Since the higher quality $I$-band data cover the same portion of the light curve, $H$-band data are not used in the derivation of model parameters. La Silla $I$-band data further confirm the last half of the second feature. Because Bronberg provides the most crucial coverage of the anomalies, we conduct three additional independent reductions of Bronberg data using DoPhot \citep{schechter93}, the DIA reduction package developed by \citet{bond01}, and the phSIS2 pipeline, based on the ISIS 2 code of
\citet{alard00}.  All three confirm the structure of the light curve in this critical region.

Note that the pronounced misalignment of the Bronberg and $\mu$FUN SMARTS data
in the middle panel is real: because $\mu$FUN SMARTS data do not cover the peak,
the $f_{\rm s}$ and $f_{\rm b}$ parameters are permitted much more freedom to match
the single-lens model than is the case for Bronberg.

The relatively low amplitude of the residuals from a single-lens model, along with the fact that these residuals are apparent over most of the duration of the source diameter crossing, generally indicate that the central caustic structure due to a companion to the lens is only magnifying a fraction of the source at one time \citep{griest98,han07}. This suggests that $w$, the ``short diameter'' or ``width'' of the central caustic is smaller than or comparable to the diameter of the source (see \citealt{chung05}). Prominent deviations from the single-lens model occur where the limb of the source enters and exits the caustic. This behavior, which is qualitatively very similar to that of MOA-2007-BLG-400 \citep{dong08}, prompts us to investigate possible two-point-mass lens (planetary or binary) models. 

\subsection{Searching a Grid of Lens Geometries
\label{sec:grid}}

The finite-source two-point-mass lens magnification calculations are carried out using the improved magnification map technique of \citet{dong06, dong08}, which is optimized for high-magnification events.  The fitting procedure follows closely that of \citet{dong08}.  The initial search for two-point-mass lens solutions is conducted over a grid of three parameters:  the short-caustic width $w$, companion/star mass ratio $q$, and the angle of the source trajectory relative to the companion/star axis $\alpha$. Since $w$ is a function of $q$ and the companion/star separation $d$, this is equivalent to also fixing $d$ at various values. The remaining parameters ($t_0$,$u_0$,$t_\e$,$\rho$) are allowed to vary. Two additional parameters for limb darkening are given fixed values, as will be discussed in Section \ref{sec:limb darkening}. The source flux $f_s$ and the blended flux $f_b$ are fit independently for each filter and telescope. We use Monte Carlo Markov Chain (MCMC) to minimize $\chi^2$ with respect to ($t_0$,$u_0$,$t_\e$,$\rho$) at each of the $(w,q,\alpha)$ grid points. There is a well known degeneracy such that for $q\ll 1$, planet/star separations $d$ and $d^{-1}$ will produce almost identical central caustic structures and consequently indistinguishable light curves for high magnification events such as this \citep{griest98}. We explore a $(w,q)$ grid for each geometry, searching the $d \geq 1$ (in units of the Einstein radius) regime for `wide' solutions and the $d<1$ regime for `close' solutions. 

\begin{deluxetable}{lllllllll}
\tablecolumns{9}
\tablecaption{MOA-2008-BLG-310 Best-Fit Planetary Model Parameters}
\tablehead{\colhead{ } & \colhead{$d$} & \colhead{$q\times 10^4$} & \colhead{$t_0$(HJD')} & \colhead{$u_0\times 10^3$} & \colhead{$t_\e$(days)} & \colhead{$\alpha$(rad)} & \colhead{$t_*$(days)} & \colhead{$\chi^2/dof$\tablenotemark{a}}}
\startdata
Wide & 1.085	& $3.31$	& 4656.39975	& $3.00$	& 11.14	& 1.21	& $0.05487$ & 2891.40/3050\\
Close & 0.927	& $3.20$	& 4656.39975	& $3.01$	& 11.08	& 1.21	& $0.05483$ & 2893.46/3050\\
Error & 0.003 & $0.26$ & \phn\phn\phn 0.00005 & $0.14$ & \phn 0.50 & 0.02 & 0.00009 & -\\
\enddata
\label{tab:params}
\tablenotetext{a}{single-lens $\chi^2/dof=3773.23/3053$}
\end{deluxetable}

\subsection{Best-Fit Model
\label{sec:best-fit}}

An initial search for two-point-mass lens solutions is conducted over the range of caustic widths $-3.5\leq\log w\leq -1.0$ (in units of the Einstein radius), companion mass ratios  $-5.0\leq\log q\leq 0$, and source trajectory angles $0\leq \alpha \leq 2\pi$ in the two separate regimes $d \geq 1$ and $d < 1$. This initial search gives us fairly good estimates of the best-fit parameters and the location of the $\chi^2$ minima in terms of $w$ and $q$ (and hence also $d$). For this particular event, however, $w$ and $q$ turn out to be highly correlated. We conduct a refined search over a grid in $(d,q)$ instead of $(w,q)$, and we also allow $\alpha$ to vary as a MCMC variable, rather than discretely. The solid black lines in Figure~\ref{fig:contours} show $\Delta\chi^2=1,4,9$ contours in the $(d,q)$ plane for the wide ({\it top}) and close ({\it bottom}) solutions, respectively. For the wide solution, the $\chi^2$ minimum occurs at $d=1.085 \pm 0.003$ and $q=(3.31\pm 0.26)\times 10^{-4}$. The close solution minimum occurs at $d=0.927 \pm 0.003$ and $q=(3.20\pm 0.26)\times 10^{-4}$. The mass ratio indicates that the companion to the lens is in fact a planet. As expected, we recover the $d\leftrightarrow d^{-1}$ degeneracy. The wide solution is favored by just $\Delta\chi^2=2.06$, indicating that the wide/close degeneracy cannot be clearly resolved in this case. The best-fit parameters for both wide and close solutions are recorded in Table~\ref{tab:params}.  Two independent algorithms were used to explore parameter space and both returned essentially identical best fits.

The wide and close planetary models qualitatively explain several features of the single-lens residuals. The lower panel of Figure~\ref{fig:caustics} shows the 
extended source at key points in time on its trajectory. The nearly identical central caustics generated by the best-fit wide and close models are both shown. The most prominent features in the residuals, shown in the upper panel of Figure~\ref{fig:caustics}, occur as the limb of the source crosses the caustic. The positive and negative spikes that are most evident from the raw data (features 2 and 4 of Fig.~\ref{fig:caustics}) coincide with the limb of the source entering the strong curved portion of the caustic and exiting the weaker straight segment. Residual patterns like this, characterized by short duration perturbations separated by a relatively flat region, are typical of planetary lens systems affected by strong finite source effects \citep{griest98,dong08,han09}. The bottom panel of Figure~\ref{fig:lightcurve} shows the residuals to the best-fit wide planetary model. The deviations from the point-lens model that initially indicated that the lens was not being accurately modeled are no longer apparent. The planetary model decreases $\chi^2$ by $\sim 880$ for three additional degrees of freedom, making this a strong detection of a Saturn mass-ratio planet/star system.

\section{Limb Darkening
\label{sec:limb darkening}}

Since the primary deviations in the light curve occur at the limb of the star, it is important to examine the effects of limb darkening on the planetary solution. For this purpose we adopt a surface brightness profile of the form
\begin{equation}
{S(\vartheta)\over S_0} = 1 
- \Gamma\biggl[1 - {3\over 2}(\cos\vartheta)\biggr]
- \Lambda\biggl[1 - {5\over 4}(\cos^{1/2}\vartheta)\biggr],
\label{eqn:ld}
\end{equation}
where $\vartheta$ is the angle between the normal to the surface of
the star and the line of sight \citep{an02}. This is somewhat more
complicated than the typically used linear limb darkening, but it is
justified by an improvement in the fit of $\Delta \chi^2 \sim
10$. We simply fix the MOA, CTIO, and La Silla limb-darkening parameters
at $(\Gamma,\Lambda) =
(0.077,0.549)$ corresponding to $(c,d)=(0.099,0.584)$ from
\citet{claret00}. These
parameters pertain to a star with $T_{\rm eff} = 5750\,$K and $\log g
= 4.0$, i.e., a post-turnoff G star, corresponding to the
$(V-I)_0=0.69$ and $M_I=3.46$ that we derive from the color-magnitude
diagram by assuming that the source suffers the same extinction and is
at the same distance as the bulge clump.  

However, because $\mu$FUN Bronberg provides the bulk of the observations
covering the peak of the event and the deviations at the limb of the
source, it is most critical that the limb darkening be accurately
modeled for these data.  We first determine the effective bandpass of
Bronberg (which is unfiltered) by making a color-color diagram of stars in
the field with colors similar to that of the microlensed source.  We
find that
\begin{equation}
\Delta(R_{\rm Bron} - I) = 0.50\,\Delta(V-I),
\label{eqn:color}
\end{equation}
i.e., almost exactly what would be expected for standard $R$ band. 
We therefore begin by adopting $(\Gamma,\Lambda) = (0.166,0.543)$ 
corresponding to
$(c,d) = (0.204,0.557)$ from \citet{claret00} for $R$ band,
As a check on this
procedure, we also allow $\Gamma$ and $\Lambda$ 
for Bronberg data to vary (along with most other parameters),
but still holding the $I$-band limb-darkening parameters fixed 
at the \citet{claret00} values for the other observatories.
The best-fit models for wide
and close planet/star separations have
$(\Gamma,\Lambda)=(-0.200,1.277)$ and $(\Gamma,
\Lambda)=(0.069,0.732)$ respectively. We find that the resulting
surface brightness profiles are similar to those defined by
the \citet{claret00} parameters.

\begin{figure}
		\includegraphics[width=1.00\textwidth]{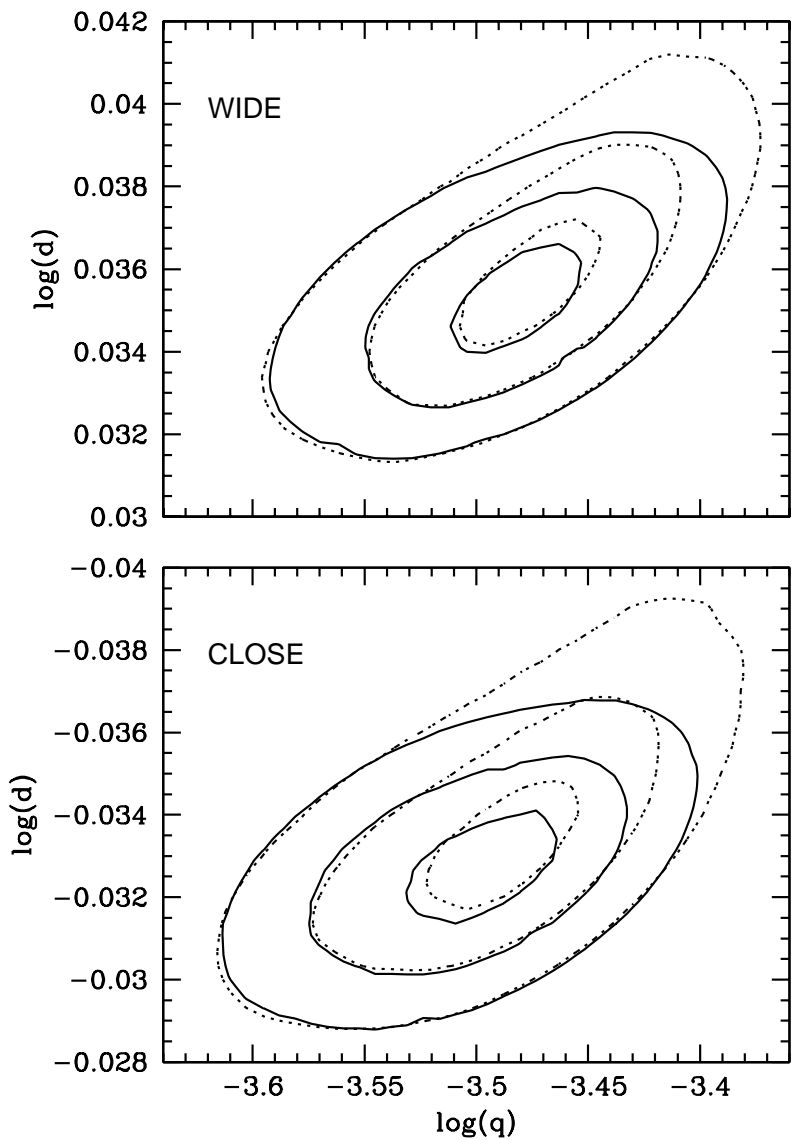}
		\caption{Top: The wide solution $\Delta\chi^2=1,4,9$ contours in the $(d,q)$ plane. The contours generated by fixing the limb darkening parameters at the \citet{claret00} values ({\it solid lines}) are similar to those from allowing $\Gamma$ and $\Lambda$ to vary freely ({\it dotted lines}). In particular, large regions of the $\Delta\chi^2=1$ minima overlap. Bottom: $\Delta\chi^2$ contours for the close solution.}
	\label{fig:contours}
\end{figure}

\begin{figure}
	\centering
		\includegraphics[width=0.9\textwidth]{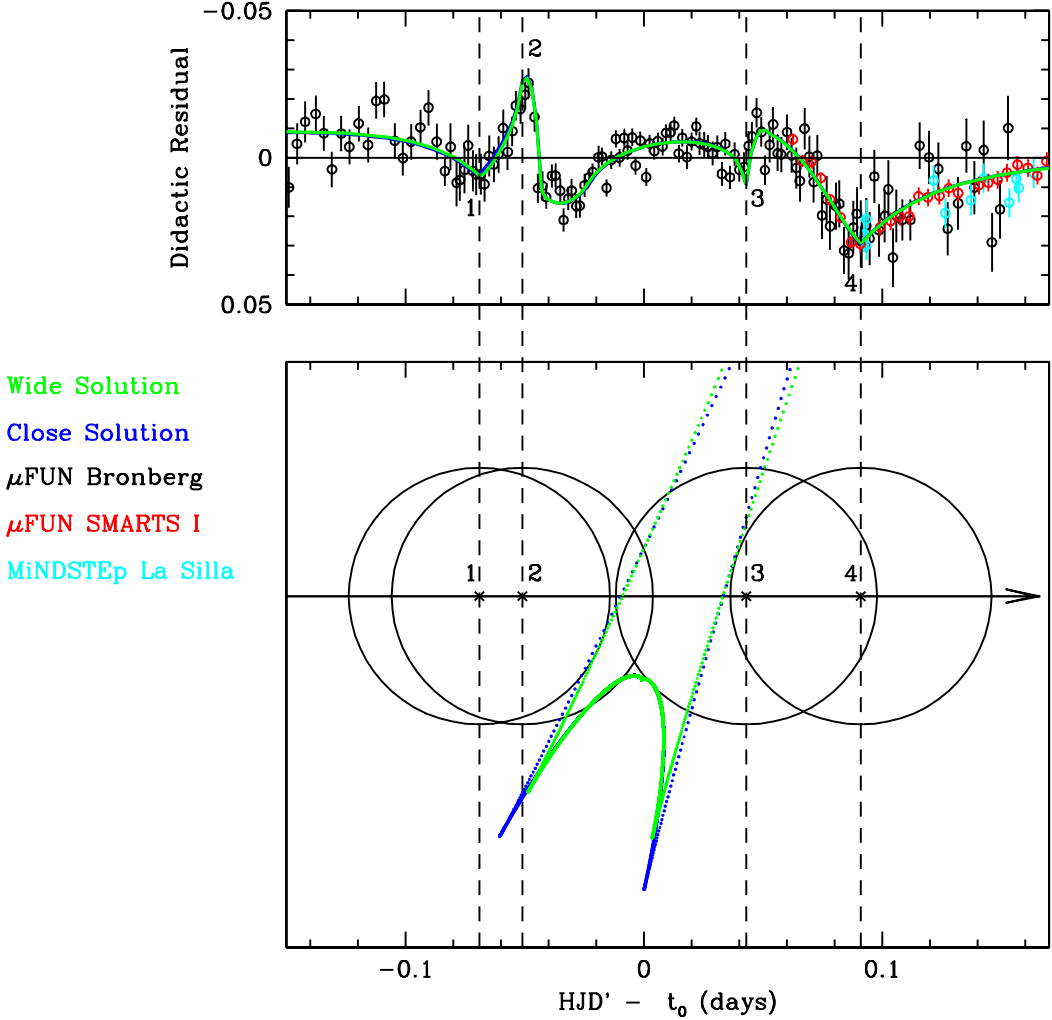}
		\caption{Top: Didactic residuals to the single-lens model. Data points are shown for $\mu$FUN Bronberg ({\it black}), $\mu$FUN SMARTS ({\it red}), and MiNDSTEp La Silla ({\it cyan}). The solid lines are the best-fit wide ({\it green}) and close ({\it blue}) planetary models. Bottom: The source trajectory ({\it solid black line}) showing the extended source ({\it circle}) crossing the caustic created by the planet at key points in time. The circle radius on the plot is the source radius crossing time, $t_* = \rho t_{\rm E} \sim 0.055$\, days. The caustics for the wide and close models are plotted in green and blue, respectively. The density of the caustic points is proportional to the strength of the caustic, so that the
``solid lines'' correspond to stronger magnification while the ``dotted lines'' indicate that the caustic is weaker. The two caustic structures are nearly indistinguishable in regions probed by the source. Several of the anomalous features apparent in the the residual plot correspond to the limb of the source crossing the caustic. These features are numbered, and dashed black lines connect them to the corresponding position of the source. Didactic residuals show the difference between the data and a point-lens model that has the same ($t_0$, $u_0$, $t_{\rm E}$, $\rho$, $f_s$, $f_b$) as the best-fit planetary model.}
	\label{fig:caustics}
\end{figure}

We further investigate the effect of limb-darkening parameters on
the final results by comparing likelihood contours for $(d,q)$ for
the two cases just described.  The
$\Delta\chi^2$ contours for the close and wide planetary models with
limb darkening fixed at the \citet{claret00} values are shown in
Figure~\ref{fig:contours} as the solid lines. These contours are
similar to the dotted lines in Figure~\ref{fig:contours} generated by
allowing the parameters for limb darkening to vary freely. Most
importantly, the best-fit values of $d$ and $q$ change by much less
than one sigma. This justifies fixing the parameters for limb
darkening at the \citet{claret00} values for all models that follow.

The temperature and metallicity we use are slightly different than those obtained by \citet{cohen09} from spectroscopy of the event. The spectroscopy yields $(\Gamma,\Lambda)=(0.105,0.535)$ ($(c,d)=(0.133,0.564)$) for $I$ band and $(\Gamma,\Lambda)=(0.203,0.517)$ ($(c,d)=(0.248,0.525)$) for $R$ band. Inserting these values into the wide and close models, we obtain best-fit parameters well within one sigma of those recorded in Table~\ref{tab:params} and contours essentially identical to those in Figure~\ref{fig:contours}.

\section{Measurement of Angular Einstein Radius $\theta_{\rm E}$
\label{sec:radius}}

The color and magnitude of the source allow us to determine its angular radius $\theta_*$, which in turn can be used to place constraints on the lens mass and lens-source relative parallax. We begin by measuring the color and magnitude of the source and the clump centroid in the calibrated CTIO field, $[(V-I),I]_{\rm source}=(1.48\pm0.01,19.28\pm0.05)$ and $[(V-I),I]_{\rm clump}=(1.84, 15.62)$. The source magnitude is derived from the microlens model and is the same for the close and wide solutions. At Galactic longitude $l=-4.08$, the angle of the bar-shaped bulge places the peak of the red clump density behind the Galactic center by 0.05 mag \citep{nishiyama05}. Assuming a distance to the Galactic center of 8\,kpc, the dereddened position of the clump is then $[(V-I)_0,I_0]_{\rm clump} = (1.05,14.37)$. Thus the extinction toward the source is $[E(V-I),A_I]=(0.79,1.25)$. We find the dereddened color and magnitude of the source, $[(V-I)_0,I_0]_{\rm source} = (0.69,18.03)$. Applying the method of \citet{yoo04}, we convert $(V-I)$ to $(V-K)$ using the color-color relations of \citet{bessell88}, and we obtain  $[(V-K)_0, K_0]_{\rm source}=(1.48, 17.24)$. We then use the color/surface-brightness relations of \citet{kervella04} to calculate the angular source radius,
\begin{equation} 
\theta_* = 0.76\pm 0.05\,\mu {\rm as}, 
\label{eqn:thetastar}
\end{equation} 
which (as with the next three equations) applies equally to both the wide and close solutions.
The source crossing time $t_*$ is 
\begin{equation}
t_* \equiv \rho t_{\rm E} = 0.05485\pm 0.00009\,\rm days,
\label{eqn:tstar}
\end{equation}
which implies that the (geocentric) proper motion, $\mu_{\rm geo} = \theta_*/t_*$, is 
\begin{equation}
\mu_{\rm geo} = 5.1\pm 0.3\,{\rm mas\,yr^{-1}}
\label{eqn:mugeo}
\end{equation}
The inferred Einstein radius, $\theta_{\rm E} = \mu_{\rm geo}t_{\rm E}$, is then
\begin{equation}
\theta_{\rm E} = 0.155\pm 0.011\,{\rm mas}\quad 
\label{eqn:einstein}
\end{equation}
The fractional uncertainties in $\theta_*$, $\theta_{\rm E}$, and $\mu_{\rm geo}$ are comparable, a typical result for point-lens events with finite source effects \citep{yee09}.
We can relate the lens mass $M_L$ to the source-lens relative parallax $\pi_\rel$ (see \citealt{gould00} for details),
\begin{equation}
M_L = {\theta_\e^2 \over \kappa \pi_\rel}
\label{eqn:mass}
\end{equation}
where $\kappa =4G/c^2\,{\rm AU}\sim 8.1\,{\rm mas}\,M_\odot^{-1}$.
If we require that $M_L>0.08\,M_\odot$ (that is, if the lens is a star) then it follows that $\pi_\rel<37\,\mu{\rm as}$.
Assuming $D_S > 8$\,kpc (as discussed above, the Galactic bar at $l\sim -4^\circ$ lies behind the Galactic center), this gives a lower limit on the distance to the lens $D_L>6$\,kpc. We conclude that if the lens mass is above the hydrogen burning limit, then it must be located in the Galactic bulge. In order to verify the bulge location of the lens, we would need another independent relation between the lens mass and distance. This could be obtained by measuring either the microlensing parallax or the flux from the lens. 

\section{Parallax
\label{sec:parallax}}

Determining the microlensing parallax $\bpi_{\rm E}$ gives us a independent relationship between the lens mass and source-lens relative parallax \citep{gould00}. The magnitude of the vector is given by, 
\begin{equation}
\pi_{\rm E}=\sqrt{\pi_\rel\over \kappa M_{\rm L}}
\label{eqn:par}
\end{equation}
while the direction is the same as that of $\bmu_{\geo}$, the lens-source relative proper motion in the geocentric frame.
In combination with the independent relation between $M_{\rm L}$ and $\pi_\rel$ obtained from the proper motion of the source,
it would be possible to give physical values to both of these parameters. With this goal in mind we examine the effects
on the light curve from two sources of parallax. Orbital parallax is caused by the acceleration of the Earth on its orbit. 
Terrestrial parallax arises from two or more widely separated observatories simultaneously observing a slightly 
different light curve due to their different vantage points. For this event orbital parallax is not expected to be detectable since the timescale is so short ($t_\e = 11.1$\,days). We expect terrestrial parallax to be poorly constrained as well. Earth-based parallax measurements require that short duration caustic crossings be observed by two or more telescopes simultaneously \citep{hardy95}. While $\mu$FUN Bronberg and $\mu$FUN SMARTS both observed the second prominent deviation as the limb of the source exited the caustic, this feature is washed out by finite source effects. We once again search the $(d,q)$ grid, allowing the north and east components of both terrestrial and orbital parallax to vary as additional MCMC parameters. We also test the case of the source-lens minimum separation $u_0\leftrightarrow -u_0$ as this is a known degeneracy in determining parallax \citep{smith03}. For the four cases ($\pm u_0$, close/wide), the reduction in $\chi^2$ ranges from 2 to 6, i.e., barely different from the $\Delta\chi^2=2$ expected from reducing the degrees of freedom by 2. The marginal detection of parallax at $\Delta\chi^2=6$ favors $\pi_{\rm E}\simeq 4$. Such a large parallax yields $\pi_\rel=\pi_{\rm E}\theta_{\rm E}=0.65\,{\rm mas}$ and lens mass $M_{\rm L}=0.005\,M_\odot$. We do not give much weight to this marginal parallax detection and the free-floating planet solution (with sub-Earth mass moon) it implies since from previous experience we have found that such small $\Delta\chi^2$ could easily be produced by low-level systematics (see also \citealt{poindexter05}). Hence we obtain essentially no new information, and, as our results are consistent with zero orbital and terrestrial parallax, we set $\pi_{\rm E}=0$ except where explicitly indicated.  

\section{Blended Light
\label{sec:blend}}

From the best-fit model we obtain a measure of how much light is being lensed in the event, in other words the flux
of the unmagnified source. In addition to the source flux, there is blended light that is not being lensed,
which may come from unrelated stars along the line of sight, companions to the source or lens, or the lens itself.
An alternate route to obtaining the lens mass and distance is possible if the flux from the lens can be isolated \citep{han05,bennett07}.

We have $H$-band images of the event taken from CTIO.
We have additional post-event $JHK$ Infrared images of MOA-2008-BLG-310 taken with the
IRSF telescope in South Africa on 2008 August 4 and the adaptive optics system NACO on the ESO VLT on 2008 July 28 under ESO Program ID 081.C-0429(A).
The pixel scales are respectively 0.27, 0.45 and 0.027\,arcsec.
A log of the VLT NACO and IRSF observations is given in Table~\ref{tab:photom}.   As detailed in Appendix~\ref{sec:calib}, IRSF serves as a bridge between
NACO and CTIO, being wider than the former and deeper than the latter.

\begin{deluxetable}{llll}
\centering
\tablecolumns{4}
\tablecaption{Log of Observations}
\tablewidth{0pt}
\tablehead{\colhead{Image} & \colhead{Date} & \colhead{hour} & \colhead{FWHM}}
\startdata
$J_{\rm IRSF}$  & 2008-08-04  & 18:05:25 & 1.4\,arcsec \\
$H_{\rm IRSF}$  & 2008-08-04  & 18:05:25 & 1.4\,arcsec \\
$K_{\rm IRSF}$  & 2008-08-04  & 18:05:25 & 1.3\,arcsec \\
\hline
$J_{\rm NACO}$  & 2008-07-28  & 01:29:20  & 0.15\,arcsec \\
$H_{\rm NACO}$  & 2008-07-28  & 02:18:29  & 0.13\,arcsec \\
$K_{\rm NACO}$  & 2008-07-28  & 00:36:17  & 0.15\,arcsec \\
\enddata
\label{tab:photom}
\end{deluxetable}

The NACO image reveals two additional stars in the vicinity of the source that are unresolved by the observations used in the light curve analysis. One of these is 3 mag brighter than the slightly magnified ($A=1.09$) source and 0.85\,arcsec away (star 3 in Fig.~\ref{fig:charts}) while the other is 0.2 mag brighter and 0.5\,arcsec away. To definitively identify the source from among this group, we create a template image from the best CTIO $I$-band images and subtract this template from 20 different astrometrically aligned, good-seeing images near the peak of the event. The magnified light of the source is isolated on the subtracted image because the contribution from other stars is removed. Thus the relative astrometry of the source is very precisely determined. DoPhot is used to find the positions of other stars on the template CTIO $I$ and median NACO $H$ images. We select 14 isolated stars common to both images and calculate the coordinate transformation from CTIO to NACO. The position of the source transformed to NACO coordinates is $13\pm 25$\,mas from the centroid of the target in Figure~\ref{fig:charts}. The nearest neighboring star is 400\,mas from the source position, and thus the identification of the source with the target on the NACO frame is very secure.    

We reduce the IRSF images following standard procedures, and measure the fluxes and positions of stars using
the DoPhot software. The reduction of the NACO images is a more complicated procedure, and is detailed in Appendix~\ref{sec:NACO}. Our goal is to put NACO photometry of the target (blend + magnified source) on the CTIO
photometric system, so that it can be compared with the source-only
$H$-band flux, which is well-measured from the CTIO $H$-band
light curve.  In principle, this could be done using
comparison stars common to NACO and CTIO $H$ band.  However, there are only
two such stars, and they have relatively large photometric errors
in CTIO photometry.  Instead, we use a large number of common stars to
photometrically align the CTIO and IRSF systems, which can therefore
be done very accurately. We then align the NACO and IRSF systems
based on 4 common stars, which have much smaller errors and 
consequently show smaller scatter than the CTIO stars. We align the IRSF system with 2MASS, allowing us to determine the calibrated magnitudes of the target and reference stars recorded in Table~\ref{tab:stars}. The photometric calibration of IRSF, NACO, and CTIO images is discussed in greater detail in Appendix~\ref{sec:calib}. We stress that in the following discussion 
(as well as the Appendices), we deliberately work in uncalibrated 
$H_{\rm CTIO}$ magnitudes, since measurement of the fraction of blended flux
depends only on relative photometry.  We apply the calibration only at
the end of this procedure to avoid introducing additional uncertainty
into the fairly subtle differential measurements.

\begin{deluxetable}{llllllll}
\tablecolumns{8}
\tablecaption{Photometric data for H CTIO, JHK IRSF, JHK NACO}
\tablehead{\colhead{Star ID}  & \colhead{$H_{\rm CTIO}$} & \colhead{$H_{\rm IRSF,calib}$} & \colhead{$H_{\rm NACO,calib}$} & \colhead{$J_{\rm IRSF,calib}$} & \colhead{$J_{\rm NACO,calib}$} & \colhead{$K_{\rm IRSF,calib}$} & \colhead{$K_{\rm NACO,calib}$}}
\startdata
1 & 16.95 &13.094 & 13.106 & 13.872 &  13.855 & 12.88 & 12.898 \\
2 & 17.69 &13.834 & 13.826 & 14.230 &  14.225 & 13.77 & 13.76  \\
3 &  -    &14.340& 14.352 & 14.83  &  14.884 & 14.24 & 14.246 \\
target & -      &  -    & 17.47  &  -     &  18.068 & -     & 17.349
\enddata
\label{tab:stars}
\end{deluxetable}

\subsection{Estimation of the target flux in $H$ CTIO}

In the following, the term ``source'' refers to the star that was microlensed, while the term ``target'' refers to all the light that is aligned with the source in the NACO images.
We calibrate the NACO $H$-band magnitude of the target via the route IRSF-to-2MASS and find
$H_{\rm NACO,calib} = 17.47 \pm 0.05$. Then using the IRSF-to-CTIO transformation, we convert the measured NACO flux into the instrumental CTIO system, $H_{\rm target, CTIO} = 21.29 \pm 0.05$. We stress that this indirect road NACO-to-IRSF-to-CTIO is actually the most accurate one to estimate the magnitude in the instrumental CTIO system.

We also carry out the following independent check. We measure aperture
fluxes $f_i$ on the NACO image for stars 1, 2, 3, and the target listed in Table~\ref{tab:stars}.
For the target, we correct the result for contaminating flux from star 3
and from another much fainter nearby star. Then using the same stars 
$i=1,2,3$ on the IRSF image, we obtain an estimate of the target magnitude on the
IRSF system: $H_{{\rm target};i,{\rm IRSF}} = H_{i,{\rm IRSF}} + 2.5\log(f_i/f_{\rm target})$.
We take the average of these three estimates (whose standard error
of the mean is only 0.012 mag), and then apply the previously derived conversion from IRSF to CTIO.
We  find $H_{\rm target, CTIO} = 21.27$.

As a further sanity check, we apply a similar procedure to compare the
NACO and CTIO images directly.  As stated at the outset, we expect that
this will be less accurate both because there are only two viable
comparison stars (1 and 2) and because the CTIO flux measurements are
less accurate than those of IRSF.  Nevertheless, we find a similar
result: $H_{\rm target, CTIO} = 21.32$, although with substantially worse precision.

We finally adopt $H_{\rm target, CTIO} = 21.28\pm 0.05$, where the error bar
reflects our estimate of the systematic error.  Clearly,
our two primary methods of estimating this quantity agree much
more closely than this, but there still could be systematic
effects common to both. We regard 0.05 mag as a conservative overestimate of the error.

Inserting the $H$-band CTIO observations into the planetary model, we
obtain the unmagnified source flux, $H_{\rm source, CTIO} = 21.55\pm
0.05$ on the instrumental CTIO system. The error bar accounts for the
the uncertainty in the fit by allowing all parameters, including
parallax, to vary freely. However, for the purpose of determining the
blend on the NACO image, we are more interested in the magnified flux
from the source at $t_{\rm NACO}$, the time the image was taken. The
magnification is determined by the separation between the source and
lens at $t_{\rm NACO}$, $u_{\rm NACO} = (t_{\rm NACO}-t_0)/t_{\rm E}$
in units of the Einstein radius. The unmagnified source flux is
anti-correlated with the Einstein crossing time $t_{\rm E}$, so the
dispersion in the magnified flux is slightly smaller than the
dispersion in the unmagnified flux. The uncertainty in the magnified
flux is related to the model uncertainty in the unmagnified flux
($f_s$) by
\begin{equation}
{{\sigma(Af_s)\over A\sigma(f_s)}} = 1 + {{d\ln A}\over{d\ln u}}
\label{eqn:diferr}
\end{equation}
which in the point-lens approximation (generally valid on the
wings on the light curve) translates to\footnote{In the limit of large A: $1 + {{d\ln A}\over{d\ln u}} =  {3\over 4 A^2}\bigl(1 + {1\over 6 A^2} + {1\over 32 A^4} + \ldots\bigr)$, which converges very quickly, even for $A\sim 2$.}
\begin{equation}
{{\sigma(Af_s)\over A\sigma(f_s)}} = 3 -{2 A^2}[1 -(1-A^{-2})^{3/2}].
\label{eqn:diferr2}
\end{equation}
In our case, the analytical result ${{\sigma(Af_s)\over A\sigma(f_s)}} = 0.77$ is very close the result calculated using MCMC data. 
 
Figure~\ref{fig:histograms} shows probability distributions for the magnified source flux constructed from the MCMC chains. Without parallax (black histogram), the close and wide solutions give the same source magnitude at the time of the NACO image, $H_{\rm magnified, CTIO} = 21.45\pm 0.04$, while the best-fit solution with unconstrained parallax (gray) gives a flux $\sim 2\%$ brighter. As noted in Section \ref{sec:parallax}, the best-fit parallax implies a planetary lens mass and is likely a spurious detection. If we constrain parallax so that the lens mass is at least $0.08\,M_\odot$ ($\pi_{\rm E}\le 0.25$), the best-fit magnified source flux is identical to the case of no parallax. Thus our best estimate of the flux strictly from the source is $H_{\rm magnified, CTIO} = 21.45\pm 0.04$, while the light aligned with the event on the NACO image is $H_{\rm target, CTIO} = 21.28\pm 0.05$. We consider $21.28 + 0.05 = 21.33$ to be a robust lower limit on the amount of light detected in the NACO image. Therefore excess light unrelated to the source is detected at the 3-sigma level. 

\section{Constraints on the Origin of the Blended Light}
\label{sec:constraints}

There are four possible causes of the excess flux detected
in the VLT NACO images: the lens, a companion to the lens,
a companion to the source, or an ambient star unrelated to
the event. 
We can estimate the probabilities of two of these possibilities (a companion to the source and an ambient star) fairly robustly using relatively well-constrained priors. However, estimates of the probabilities of the other two possibilites depend on relatively poorly-constrained assumptions about the distribution of planets orbiting hosts of various masses, including low-mass stars and brown dwarfs.  Therefore, the relative probabilities of these four possibilities also depend on these assumptions, and so we cannot robustly distinguish between them.  We will therefore discuss the constraints we have on each possibility in turn, but will not attempt to assess which is the most likely interpretation.  Rather, we will simply review the observations that are required to distinguish between these possibilities in Section \ref{sec:characterize2}.

One constraint that applies to all four possibilities
is that the excess light must lie within 1 FWHM of the source.
We find, by adding 17\% of the target flux at various positions,
that separations greater than 1 FWHM would permit resolution
of the excess flux.  The PSF of the VLT NACO image is slightly elongated:
($140\,{\rm mas}\times 124\,{\rm mas}$), so we adopt 132 mas
for present purposes.

\subsection{Ambient Star}

 From direct examination of the NACO $H$ image, we find the density
of stars within 0.5 mag of the $H$-band magnitude of the
excess light to be $0.94\,\rm arcsec^{-2}$.
Hence the prior probability that such a random star
lies buried under the NACO image is 5.1\%.

\subsection{Companion to the Source}

The source is a main-sequence G star.  From Table 7 of \citet{duquennoy91},
we find that 9.4\% of their sample (of 164 stars) have companions
within the mass-ratio range of $0.57<q< 0.76$, i.e., the mass
range corresponding the $\pm 0.5\,$mag of the observed excess flux.
However, from Figure 5 of \citet{duquennoy91}, 22\% of companions
lie outside $\sim 1000$ AU, the size of the NACO PSF projected on the
source plane.  A further 3\% have orbits shorter than $\sim 3$ days, which
would have given rise to observable ``xallarap'' signals in the
microlensing light curve.  Hence, if bulge G stars are like the
local sample,  $0.094\times 0.75= 7.1\%$ of them have companions
within 0.5 mag of the observed excess flux, and lie at separations
where they would not have been detected.  This is comparable to
the corresponding value for ambient stars.

{\subsection{Companion to the Lens}
\label{sec:comp_lens}}

If the lens had a companion, it would induce shear on the
lens's gravitational field, which would generate
a small Chang-Refsdal (1979, 1984) caustic at the center
of magnification of the lensing system.  This would in turn
produce spikes in the light curve
at HJD$'$ 4656.36 and 4656.44, when the lens center-of-magnification
crosses the limb of the source.  The residuals to the planetary model
(Fig.~\ref{fig:lightcurve})
strongly limit any such spikes.  To put this constraint on
a quantitative basis, we fit the light curve to models that
have two additional parameters, $\phi$, the angle between the 
planet axis and the binary-companion axis, and $w_{\rm com}$,  
the width of the caustic induced by the companion,
\begin{equation}
w_{\rm com} = 4 {q_{\rm com}\over d_{\rm com}^2},
\label{eqn:wcomp}
\end{equation}
where $(q_{\rm com},d_{\rm com})$ are the mass ratio and separation
of the companion.  We hold $(w_{\rm com},\phi)$ at a grid of
fixed values and minimize $\chi^2$ with respect to all other
parameters.\footnote{In practice, $q_{\rm com}$ is held fixed
at 1, since we are probing the Chang-Refsdal (1979, 1984) limit,
for which all caustics with the same $w$ but various $q$ are 
essentially the same.}

This search reveals an improvement of $\Delta\chi^2= -7.3$ for
two additional degrees of freedom, with a best fit of 
$(\log w_{\rm com},\phi) = (-3.28,40^\circ)$. This improvement is
too small to claim a detection, since it could occur by chance
with probability 2.6\%, and could also be due to low-level systematics.
Thus, while this test raises the tantalizing possibility that the
excess light is due to a companion to the lens, we mainly focus on
the $3\,\sigma$ upper limits to shear from a companion with $w_{\rm com}
< 1 \times 10^{-3}$ over
almost all angles.  See Figure \ref{fig:shear}.  By comparison,
$w_{\rm planet} = 5\times 10^{-3}$ (see Fig.~\ref{fig:caustics}).

Equation (\ref{eqn:wcomp})  can also be written
$$
w_{\rm com} = {4\kappa M_{\rm com} 
\pi_{\rm rel}\over \Delta\theta_{\rm com}^2}
$$
where $M_{\rm com}$ is the companion mass, and $\Delta\theta_{\rm com}$
is its angular separation.  If the lens is in the bulge and the
excess light is due to its companion, then 
(see Section \ref{sec:mass_estimate})
$M_{\rm com}\sim 0.6\,M_\odot$, and so
$w_{\rm com}<1\times 10^{-3}$ implies
\begin{equation}
\Delta\theta_{\rm com} > 16\,{\rm mas}{\pi_{\rm rel}\over 37\,\rm \mu as}
= 16\,{\rm mas}\biggl({M\over 0.08\,M_\odot}\biggr)^{-1}.
\label{eqn:complimits}
\end{equation}
For foreground-disk lenses, the limit on $\Delta \theta_{\rm com}$
continues to grow, but much
more slowly, to $70\,$mas for $(M,\pi_{\rm rel})=
(0.003\,M_\odot,1\,\rm mas)$.

Hence, in contrast to source companions, which are permitted
over 4 decades of separation, companions to the lens are
restricted to 1--2 decades for bulge lenses and a somewhat narrower
range for disk lenses.

{\subsection{Lens Mass Estimate}
\label{sec:mass_estimate}}
	
We stress that the excess light could be due to any of the three options
above.
The excess light could also be due to the lens.  As we show below, this requires the lens star to be relatively massive and so quite close to the source.  Such small source-lens distances are generally disfavored by phase space and kinematical factors.  However, as mentioned previously, evaluating the prior probability of this scenario requires adopting a specific assumption of frequency of planets as a function of host mass, which is poorly constrained.

Under the assumption that the NACO blended light is due to the
lens, we can estimate the lens mass using its inferred instrumental
flux $H_{\rm lens,CTIO}= 23.38\pm 0.41$, the measured Einstein radius
$\theta_{\rm E}\equiv \sqrt{\kappa M_{\rm L} \pi_{\rm rel}} = 0.155\pm 0.011$,
and an assumed range of mass-luminosity relations, which depend
on the lens's unknown metallicity. While our actual calculation 
is fully self-consistent, the basic result can be understood
intuitively as follows. For any possible mass consistent with
the observed flux, the lens-source distance will be quite small,
$D_{\rm LS}\sim \pi_{\rm rel}D_{\rm S}^2/{\rm AU} =
300\,{\rm pc}\,(M_{\rm L}/0.7\,M_\odot)^{-1}$ relative to $D_{\rm S}$,
and the range of possible values due to different candidate
masses is even smaller. Hence, we can just adopt 
$D_{\rm LS} = 300\,{\rm pc}$. For fixed $D_{\rm LS}$ the lens
lies, on average, $D_{\rm LS}/2$ or 0.04 mag in front of the
local bulge density peak (defined by the clump giants).
At Galactic latitude $b=-4.56$, the dispersion of distance moduli
of bulge sources is $(5/\ln 10)\sin b/0.6 = 0.29$ mag for an
adopted bulge flattening of 0.6. We adopt an absolute magnitude
for the clump of $M_{H,\rm clump} = -1.41\pm 0.05$, and so from the
observed $H_{\rm clump,CTIO} = 17.35$, we infer
$$
M_{H,\rm lens} = M_{H,\rm clump} + 
(H_{\rm lens,CTIO} - H_{\rm clump,CTIO})
+ (0.04 \pm 0.29) = 4.66 \pm 0.50
$$
Then using six isochrones generated by the Dartmouth Stellar Evolution Database \citep{dotter08} with [Fe/H] ranging from $-0.5$ to $0.5$ and age ranging from 5\,Gyr to 10\,Gyr, we estimate the lens mass to be $M_{\rm L} = 0.67\pm0.14\,M_\odot$.

Assuming the excess light is indeed due to the lens and using the resulting estimate of the lens mass, we can now estimate the properties of the planetary companion. The planet mass is $m_p = 74 \pm 17 M_\oplus$, roughly $80\%$ the mass of Saturn.  Taking account of the uncertainties in both the distance to the lens and the angular Einstein radius, as well as the wide/close degeneracy, the projected separation between the planet and host star is $1.25\pm0.10$\,AU.

Note that if the blended light is not due to the lens, then the lens
must be fainter than this light and so (unless it is a remnant) also
of lower mass.  The planet would then be of proportionately lower mass
as well.

\section{Discussion
\label{sec:discussion}}

\subsection{Sub-Saturn Mass Planet -- Candidate Bulge Planet
\label{sec:subsaturn}}

Microlensing event MOA-2008-BLG-310 is one of only two published high
magnification events to date for which the source is as large or
larger than the central caustic. It bears many
similarities to the other, MOA-2007-BLG-400 \citep{dong08}. Like that earlier
event, the planetary perturbations in the light curve are not
immediately apparent, having been smoothed out by finite-source
effects. We find that a Saturn mass ratio planet/star model is
nevertheless favored over the single-lens model by a significant
reduction in $\chi^2$ ($\Delta\chi^2 \sim 880$ for an additional three
degrees of freedom). Using VLT NACO (together with IRSF) photometry, we
definitively detect excess light blended with the source that is due
to the lens, or a companion to the lens or the source,
or an unassociated star.  
Regardless of the origin of this excess light, however, it places
an upper limit on the lens flux and so on its mass.  The planet's
Saturn-like mass ratio therefore implies that it has a sub-Saturn mass.

Our measurement of the angular Einstein radius $\theta_{\rm E}$
constrains a combination of the lens mass and distance. We thereby
conclude that if the lens is a star, then it must be in the bulge.

We are not able to resolve the close/wide degeneracy in the
geometry of the planetary system. However the separate solutions for
the lens/star separation $d$ differ only by a factor of 1.17. The
$d\leftrightarrow 1/d$ degeneracy is not as severe in this case
because the planet is located very close to the Einstein radius.

Figure \ref{fig:detections} shows the mass versus equilibrium
temperature for planets that have been detected orbiting main-sequence
stars via radial velocity, transits, direct imaging, astrometry, and
microlensing.  The position of MOA-2008-BLG-310Lb is shown under
two assumptions.  The hexagon symbol indicates its position assuming
that the excess flux is due to the lens. We then obtain a host mass
of $M_L=0.67 \pm 0.14 M_\odot$, and so a planet mass of 
$0.23\pm 0.05\,M_{\rm jup}$.  The ``tail'' extending toward lower
masses and colder temperatures assumes that the excess light is 
not from the lens (and so is due to a companion to the source or lens).
The path of this tail is determined by the measurement of
$\theta_{\rm E} = 0.155\,$mas, which constrains the product of
the lens mass and lens-source relative parallax to be
$\theta_{\rm E}^2/\kappa = M\pi_{\rm rel} = 3\,M_\odot\,\mu$as.

\subsection{Importance of Further Characterizing the Planet
\label{sec:characterize1}}

All possibilities for the lens identification are interesting.  
In particular, the host might be a brown dwarf or free-floating planet
in the foreground disk.  In this case, the Saturn-mass-ratio
companion would actually be a moon.  Future observations with
the {\it Hubble Space Telescope (HST)} or adaptive optics (AO)
could distinguish among these three possibilities.  First, of
course, mere detection of the light from the host would confirm
it is a star and therefore that this is indeed a bulge planetary system, 
the first unambiguous such detection.

If the excess light is due to the lens, then Figure \ref{fig:detections}
shows that this detection is probing a new part of parameter space.
The previous eight planets detected via microlensing
span a relatively wide range of mass from a few Earth masses to
several Jupiter masses, but are largely located at relatively cold
equilibrium temperatures of $\sim 40-80~{\rm K}$, similar to the outer
planets of our solar system. In contrast, the radial velocity and
transit methods are generally only sensitive to sub-Saturn mass
planets with relatively warm equilibrium temperatures of $\ga 300~{\rm
K}$. Therefore, little is currently known about the demographics of
sub-Saturn mass planets with equilibrium temperatures between
$100-300~{\rm K}$. MOA-2008-BLG-310Lb
would make this the first microlensing planet to fall on the Snow Line.

Of course, if the excess light is not due to the lens, then we
have no hard information on the lens mass. However, the ``tail'' in
Figure \ref{fig:detections} indicates an interesting possibility:
that MOA-2008-BLG-310LB is a ``cold Neptune'' orbiting a low
mass star, similar to several other such detections.

\subsection{Planet Characterization using {\it HST} or AO
\label{sec:characterize2}}

What are the prospects for characterizing the host and its planet? As we
discussed in Section \ref{sec:parallax}, the event contains essentially no
parallax information. Hence, the only path toward measuring the lens
mass and distance
is direct detection of the host (or possibly its companion). 
For either the wide
or close solution, the geocentric proper motion is $\mu_{\rm
geo}=\theta_{\rm E}/t_{\rm E} = 5.1\,{\rm mas\,yr^{-1}}$. The
heliocentric and geocentric proper motions differ by
\begin{equation}
|\bmu_{\rm hel} - \bmu_{\rm geo}| = |{\bv}_{\oplus,\perp}|\pi_{\rm rel}
= {\theta_{\rm E}^2 v_{\oplus,\perp}\over \kappa M} =
0.018\,{M_\odot\over M}\,{\rm mas\,yr^{-1}}
\label{eqn:pmdif}
\end{equation}
where $v_{\oplus,\perp}=28\,\kms$ is the velocity of the Earth
projected on the plane of the sky at the peak of the event. Hence, if
the lens is luminous ($M\ga 0.08\,M_\odot$), then the heliocentric and
geocentric proper motions are essentially identical, and so the
magnitude of the heliocentric lens-source relative proper motion is
well determined.  

This known proper motion can then serve as an anchor point for the 
interpretation of future high-resolution images, which could 
in principle directly detect the lens or demonstrate unequivocally
that it is not luminous and so is a sub-stellar object in the foreground
disk.  However, as we now show, such unambiguous results actually
require that new images, with FWHM$\sim 50$ mas, be obtained ``immediately'',
i.e., before the lens and source have separated significantly.

Suppose, by contrast, the first epoch consisted solely of the 
132 mas FWHM images already in hand, and that second epoch
AO images were obtained 10 years later, which (unlike the first
epoch) did reach the diffraction limit of 50 mas.  The lens will
then be 50 mas from the source, and so separately resolved if
it is luminous.  But if such a star were observed at 50 mas,
how could we be certain it was the lens?  If the excess light
were due to an ambient star or to a companion to the lens or
the source,
then this object could also happen to be 50 mas from the source
at this latter epoch.  For the ambient-star and source-companion
cases this is obvious.
For the lens-companion case, the shear limits discussed in
Section \ref{sec:comp_lens} do place some constraints on future
lens-companion positions, but as we will make clear further below,
these still allow it to be 50 mas from the source after 10 years.

On the other hand, suppose that nothing was detected in such
10-year post-event images.  In this case, we would know that
the lens was not in the bulge, but we would still not be able
to determine whether the excess light had been due to an ambient
star, or companions to the lens or source.  And it would be of
substantial interest to do so because, in this case, a lens companion
would be the only clue to the distance (and so mass) of the lens.

Let us consider now how the situation would change if new
images were obtained immediately with 50 mas resolution, either
from {\it HST} or using AO.  Such images would either resolve out
the excess flux, or restrict it to 50 mas radius (smaller
in the case of {\it HST} as discussed below).
If it were resolved, then the appearance of a ``new'' star 10 years
later would have to be due to the lens or its companion.  (In principle,
such a ``new'' star could be an ambient star that had been hidden
at the time of the first epoch, but the probability of this
is reduced by $(50/132)^2 =0.14$ and is further reduced by the
chance that it would happen to be very close to 50 mas from the
source at the second epoch.)\ \  The proper motion of the excess
light relative to the source would tell us whether it was an ambient
star, a companion to the source or lens, which in the last case
would give the direction of proper motion, thereby confirming
that the ``new'' star was either the lens or a second (and very close) 
companion to the lens.  These last two possibilities could not be 
strictly differentiated.  However, as discussed in Section 
\ref{sec:comp_lens}, companions cannot be too close because of
the limits on shear, so the second companion could
potentially be strictly ruled out depending on the analysis
of the other stars in the image.

On the other hand, if the first epoch image did not resolve out
the excess flux, then, as argued above, the ambient star
hypothesis would be so much less likely that it could be ignored.
Appearance of a ``new'' star in the second epoch would then be
either the lens or a very close companion to the lens.  Again,
the strong constraints on the shear would translate into very
strong constraints on the lens-companion scenario.

If the first epoch were carried out with {\it HST} then these
constraints could be tightened further.  Color-dependent centroid 
shifts (between say $V$ and $I$) can be detected for star separations
down to about 15 mas (assuming an $I$-band flux ratio of 11\%),
which (as outlined in Section \ref{sec:comp_lens}) is quite close 
to the minimum source-lens separation, unless the source is in bulge.

In brief, immediate observations with {\it HST}, with followup
3 ({\it HST}) to 10 (AO) years hence, could 
unambiguously distinguish between a bulge and disk lens, and
if the former, give a good measurement of the lens mass and distance.
Immediate AO observations, if they achieved 50 mas, would
significantly constrain the possible options, but would not yield
an absolutely air-tight case.

We note that the calibrated source magnitude is $(V,I,H)_{\rm source}=(20.76,19.28,17.73)\pm 0.05$ and the $H$-band magnitude of the blend is $H_{\rm blend}=19.56\pm 0.41$. If the blend is in the bulge, then  $(V,I)_{\rm blend}\sim (24.0,21.9)$.

\subsection{Other Bulge Planet Candidates
\label{sec:candidates}}

The procedures just outlined are challenging but alternative routes to secure detection of bulge planets are, if anything, more difficult. \citet{gaudi00} discussed the prospects for detecting transiting planets in the bulge and \citet{sahu06} reported the detection of 16 candidate bulge planets from a transit survey carried out with the {\it Hubble Space Telescope}. Two of these were bright enough for radial-velocity follow-up, one of which showed variations consistent with a planet with mass $m=10\,M_{\rm Jupiter}$ and the other showed upper limits $m<4\,M_{\rm Jupiter}$. The stars are so bright, however, that their inferred masses indicate that at least one (and possibly both) probably lie in the foreground disk. Nevertheless, this technique could in principle be pushed harder, particularly when larger telescopes come on line. Even then, however, lower-mass planets, $m\la M_{\rm Jupiter}$ will probably only be accessible with microlensing.

There are three other planets detected by microlensing for which
the distances are neither measured nor strongly constrained,
OGLE-2005-BLG-169Lb \citep{gould06},
OGLE-2005-BLG-390Lb \citep{beaulieu06}, and
MOA-2007-BLG-400Lb \citep{dong08}. In addition, the distance to
MOA-2003-BLG-53/OGLE-2003-BLG-235 is not precisely enough measured to
determine unambiguously whether it is in the inner disk or the outer bulge.
In all four cases, both
$\theta_{\rm E}$ and $\mu$ are measured, so we estimate the
minimum lens mass that would allow the lens-planet system
to be in the bulge and the time that must elapse before
definitive imaging observations can be undertaken. As mentioned below, all of these events have large proper motions, $\mu \gtrsim 7\,{\rm mas\,yr^{-1}}$, which generally favor disk lenses.

For OGLE-2005-BLG-169Lb, $\theta_{\rm E} = 1.00\pm 0.22\,$mas
and $\mu = 7 - 10 \,{\rm mas\,yr^{-1}}$. Even adopting the $1\,\sigma$
lower limit on $\theta_{\rm E}$, then $\pi_{\rm rel}> 75\,\mu$as
for stellar hosts with $M\geq M_\odot$. Thus, for bulge sources
at $D_{\rm s}= 8\,$kpc, the lens distance is no more than 5 kpc.
Hence, the lens is almost certainly in the disk. Measurements to
confirm this relatively secure conclusion could be made as early
as 7 years after the event, i.e., 2012.

For MOA-2007-BLG-400Lb, $\theta_{\rm E} = 0.32\pm 0.02\,$mas
and $\mu = 8.2\pm 0.5\,{\rm mas\, yr^{-1}}$. Adopting
$D_{\rm s}= 8\,$kpc, the lens would only lie within 2 kpc of the source
provided that $M\ga 0.30\,M_\odot$. Thus, this is a reasonable,
but not particularly strong candidate for a bulge lens. The source
is a moderately bright subgiant, so for 10m class telescopes
it is perhaps best to wait for the separation to reach 70 mas,
which will require about 9 years, i.e., in 2016.

For OGLE-2005-BLG-390Lb, $\theta_{\rm E} = 0.21\pm 0.03\,$mas
and $\mu = 6.8\pm 1.0\,{\rm mas\,yr^{-1}}$. It is therefore the
best previous candidate for a bulge lens since $\theta_{\rm E}^2$,
which is the product of the mass and relative parallax, is only
a factor 1.6 times larger than for MOA-2008-BLG-310Lb. This means
that if it were at the bottom of the main-sequence, it would lie
about 3 kpc in front of the source and therefore most likely lie
in the disk, but if it had significantly larger mass it would be
in the bulge. However, in this case, the source is a G4 III giant with $I_0=14.25$,
which implies $M_H \sim -0.85$. A lens close to the bottom of the
main sequence has $M_H\sim 11$ and so (even accounting for its
closer distance) would appear 25000 times fainter than the source.
While this is an extreme case, it would appear prudent to wait for
the lens to move 3 FWHM away from the source, which for 10m
class telescopes would require about 20 years, i.e. 2025. If larger
telescopes with AO come on line before that, it will of course
be possible to make the measurement sooner.

Finally, MOA-2003-BLG-53/OGLE-2003-BLG-235 has a proper motion of
$3.3\pm 0.4\,\rm mas\,yr^{-1}$, which was sufficient to measure
the color-dependent centroid shift from {\it HST} observations taken just
1.78 years after the event, but only at the $\sim 3\,\sigma$ level
\citep{bennett235}.  Based on this measurement, the lens distance
is estimated to be $D_{\rm L} = 5.8^{+0.6}_{-0.7}$\,kpc, which reflects
a roughly 30\% error in the lens-source relative parallax
$\pi_{\rm rel}$.  Thus
this planetary system could be in the inner disk or the outer
bulge.  The color-dependent centroid shift could certainly be measured
more accurately today, but this would not dramatically decrease
the uncertainty in $D_{\rm L}$, which is fundamentally limited
by the 25\% uncertainty in $\theta_{\rm E}^2 = \kappa M\pi_{\rm rel}$,
and so in $\pi_{\rm rel}$.  Thus, pending spectra of this
$I\sim 21$ lens after it is fully separated from the source,
it will be difficult to prove whether or not this planet is
in the bulge.

\acknowledgments

We acknowledge the following support:
NASA NNX06AF40G (JJ, AG, RP, SG);
NSF AST-0757888 (SD, AG);
The Creative Research Initiative program (2009-008561)
of the Korea Science and Engineering Foundation (CH).
We thank the VLT Science Operations team for
successfully executing our NACO observations in Target of Opportunity mode.
We acknowledge the support by ``Agence Nationale de la Recherche'', the
HOLMES project, ANR-06-BLAN-0416.
Part of the data were obtained at ESO's La Silla Paranal
Observatory in Chile.
This work was supported in part by an allocation of computing time
from the Ohio Supercomputer Center. 

\appendix
\section{Reduction of VLT NACO Images}
\label{sec:NACO}

Since the reduction of the NACO images is a delicate procedure, we present it in more detail.
The master darks are median stacked from 5 raw dark frames taken on the same night with the same
integration time (40\,s for $H$ band, 50\,s for $J$ and $Ks$) as the science frames. The master flatfield is
obtained from 6 lampflats taken the same night. A badpixel map for correction of the raw frames is
obtained  using the deadpix routine from the ESO ECLIPSE package \citep{devillard97}. The science frames
(24\,s in $J$, $H$ and 49\,s in $Ks$) are then dark subtracted,  flatfielded, median co-added and sky-subtracted using the JITTER infrared data reduction software \citep{devillard99}. To avoid border effects, we keep only the intersection of frames for all the dithered positions for our photometric analysis.

We use the Starfinder \citep{diolaiti00} tool to extract the photometry of the reduced NACO frames.
Starfinder has been specially designed to perform photometry of AO
images of crowded fields.  It creates a numerical
PSF template from chosen stars within frame, which is then used for
PSF-fitting of all stars in the field. Even though the
AO correction for the given data set is good (strehl ratios of around
10\%)  and the variation of the PSF-shape across the field of view is small,
we decide to take star 3 (see Fig.~\ref{fig:charts}) as PSF template for best photometric
accuracy on the target, as it is the closest high signal-to-noise ratio star to the microlens.

\section{Photometric calibration of IRSF, VLT NACO and H CTIO}
\label{sec:calib}

As discussed in Section \ref{sec:blend}, there are only two common stars, both with relatively large photometric errors, with which to perform a direct photometric alignment between the CTIO and NACO systems. As the IRSF images share more common stars with both NACO and CTIO, we obtain a more accurate alignment using the indirect transformation NACO-to-IRSF-to-CTIO. Specifically, we perform the following steps. First, the IRSF images are calibrated with respect to 2MASS reference stars using GAIA/Skycat Fit to obtain initial star positions relative to the 2MASS astrometric catalog, and then Tweak is used to refine them. We cross identify 1521 objects between the 2MASS and IRSF frames, 779 of which have high quality flags (labeled AAA in 2MASS catalog), and then apply two further restrictions: keeping only the bright end of the sample, and
removing 1.5 sigma outliers. We adopt the color terms as given by the IRSF manual and detailed in \citet{kato07},
and we fit the zero point :

\noindent $J_{\rm IRSF,inst} = 23.073 \pm 0.001 + J_{\rm 2MASS} - 0.043 (J_{\rm 2MASS}-H_{\rm 2MASS}) + 0.018$\\
\noindent $H_{\rm IRSF,inst} = 23.128 \pm 0.001 + H_{\rm 2MASS} + 0.015 (J_{\rm 2MASS}-H_{\rm 2MASS}) + 0.024$\\
\noindent $K_{\rm IRSF,inst} = 22.334 \pm 0.001 + K_{\rm 2MASS} + 0.010 (J_{\rm 2MASS}-K_{\rm 2MASS}) + 0.014$\\

We apply these relations to the 3006 objects with good cross ID in IRSF images. Up to this point, we have calibrated
$JHK$ measurements taken by the IRSF telescope. The WCS positions of IRSF objects are deduced using the
WCSTools routine xy2sky and used as references to calibrate NACO images with the WCSTools routine imwcs. In the NACO field, we identify 6 bright stars likely not to be affected by blending when comparing IRSF and NACO. Two of them are variable, which leaves us with four stars with the color range $(J-H) = 0.4-0.78$. We note that there is no color term in the transformation, and we estimate photometric offset between $H_{\rm IRSF,calib}$ and instrumental NACO to be $27.873 \pm 0.014$ in $H$.

We cross-identify 209 stars in the IRSF and CTIO $H$-band images with matches better than $0.8''$.
We clip at $\pm 0.1$ mag around the mean of $H_{\rm CTIO}-H_{\rm IRSF,calib}$, and keep 175 stars. We estimate the zero point
offset between instrumental $H_{\rm CTIO}$ and $H_{\rm IRSF,calib}$ to be $3.8164 \pm 0.0034$.

\begin{figure}
	\centering
		\includegraphics[width=1.00\textwidth]{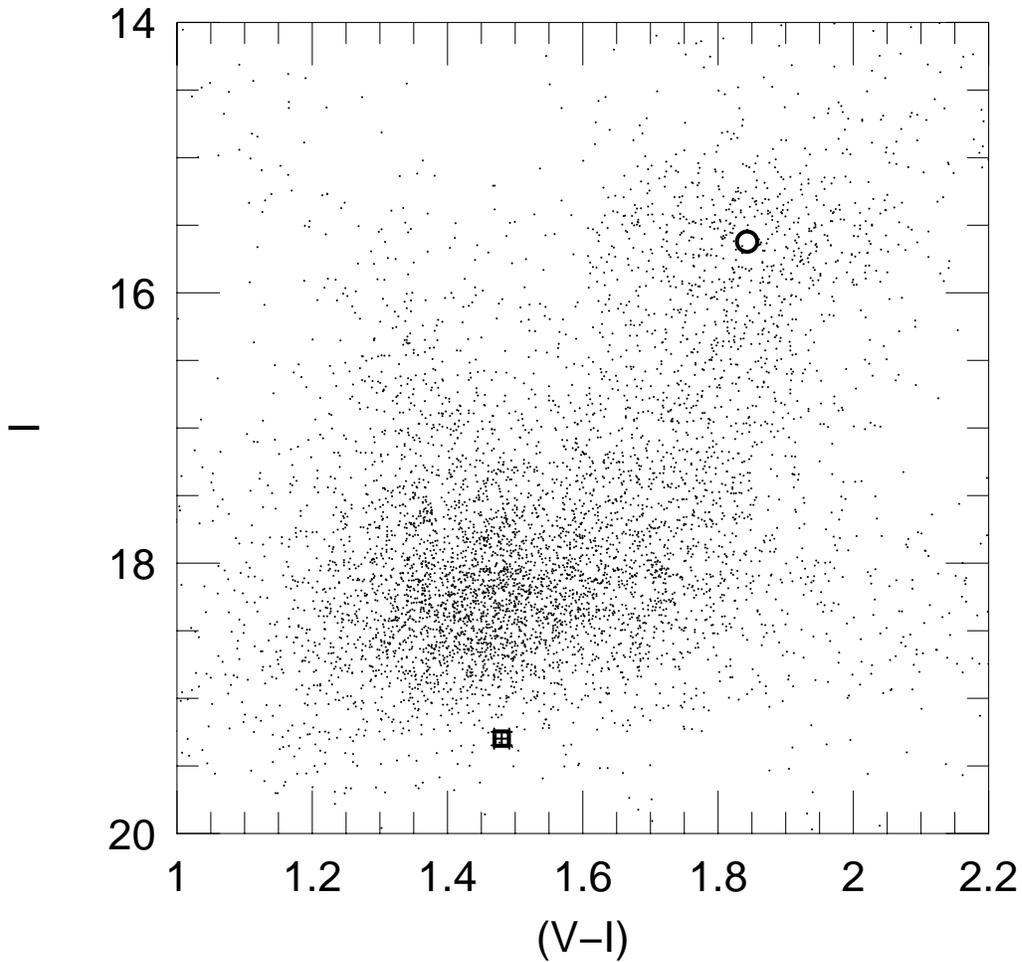}
		\caption{Calibrated color-magnitude diagram of the field containing MOA-2008-BLG-310. The clump centroid ({\it circle}) is located at $[(V-I),I]_{\rm clump}=(1.84,15.62)$. The source color and magnitude ({\it square}) is derived from the best-fit planetary model, $[(V-I),I]_{\rm source}=(1.48\pm 0.01,19.28\pm 0.05)$. Assuming the source lies at 0.05 mag behind the Galactic center, $(V-I)_0=0.69$ and $M_I=3.46$, consistent with a post-turnoff G type star, as confirmed spectroscopically by \citet{cohen09}.}  
	\label{fig:cmd}
\end{figure}

\begin{figure}
	\includegraphics[width=1.00\textwidth]{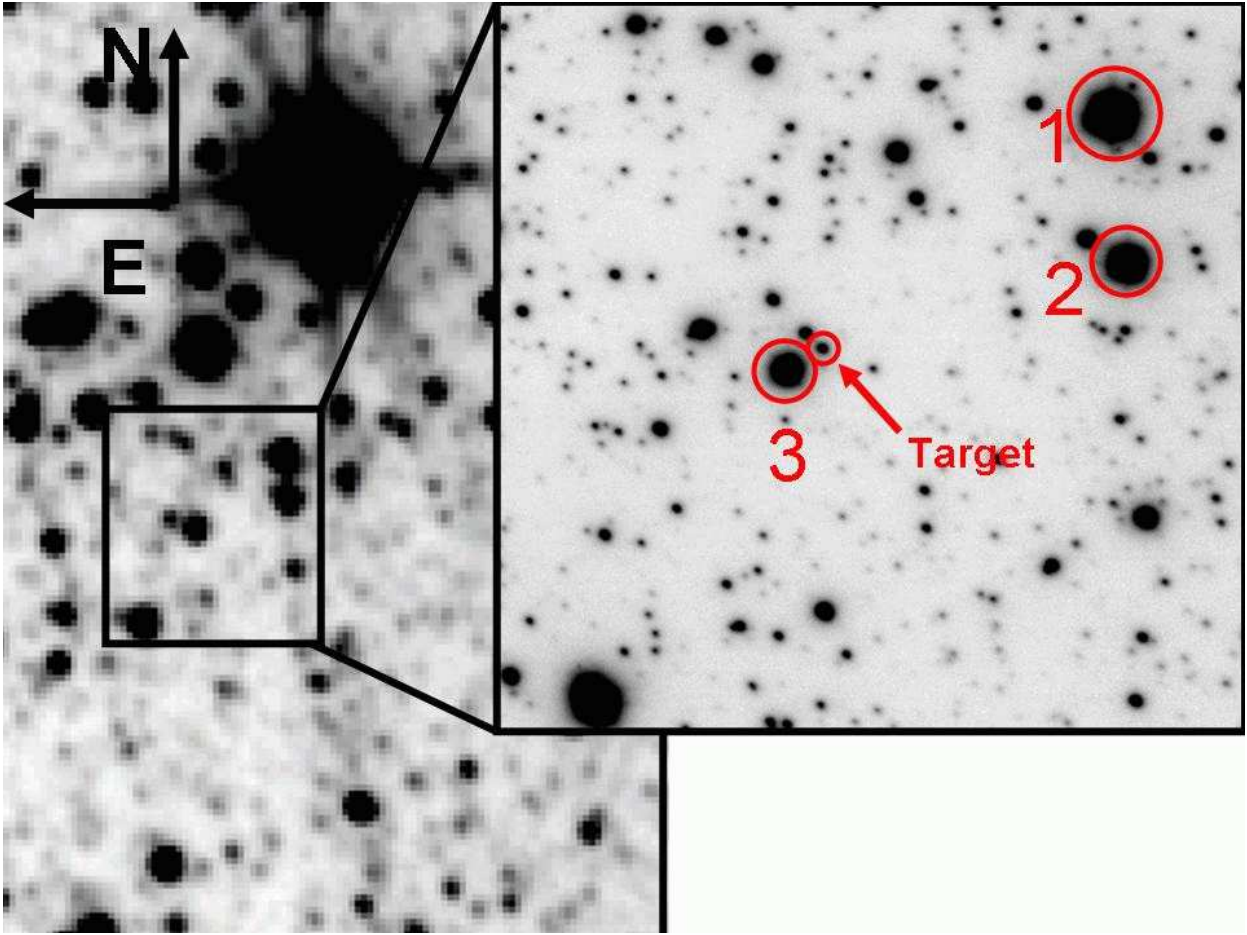}
	\caption{Left: Image taken in $H$ band by the IRSF telescope in South Africa on 2008 August 4. Right: Median $H$-band AO image taken by NACO on VLT on 2008 July 28, near the baseline of the event. The lensed source plus blend is indicated as the target and reference stars are circled and numbered.}  
	\label{fig:charts}
\end{figure}

\begin{figure}
	\includegraphics[width=1.0 \textwidth]{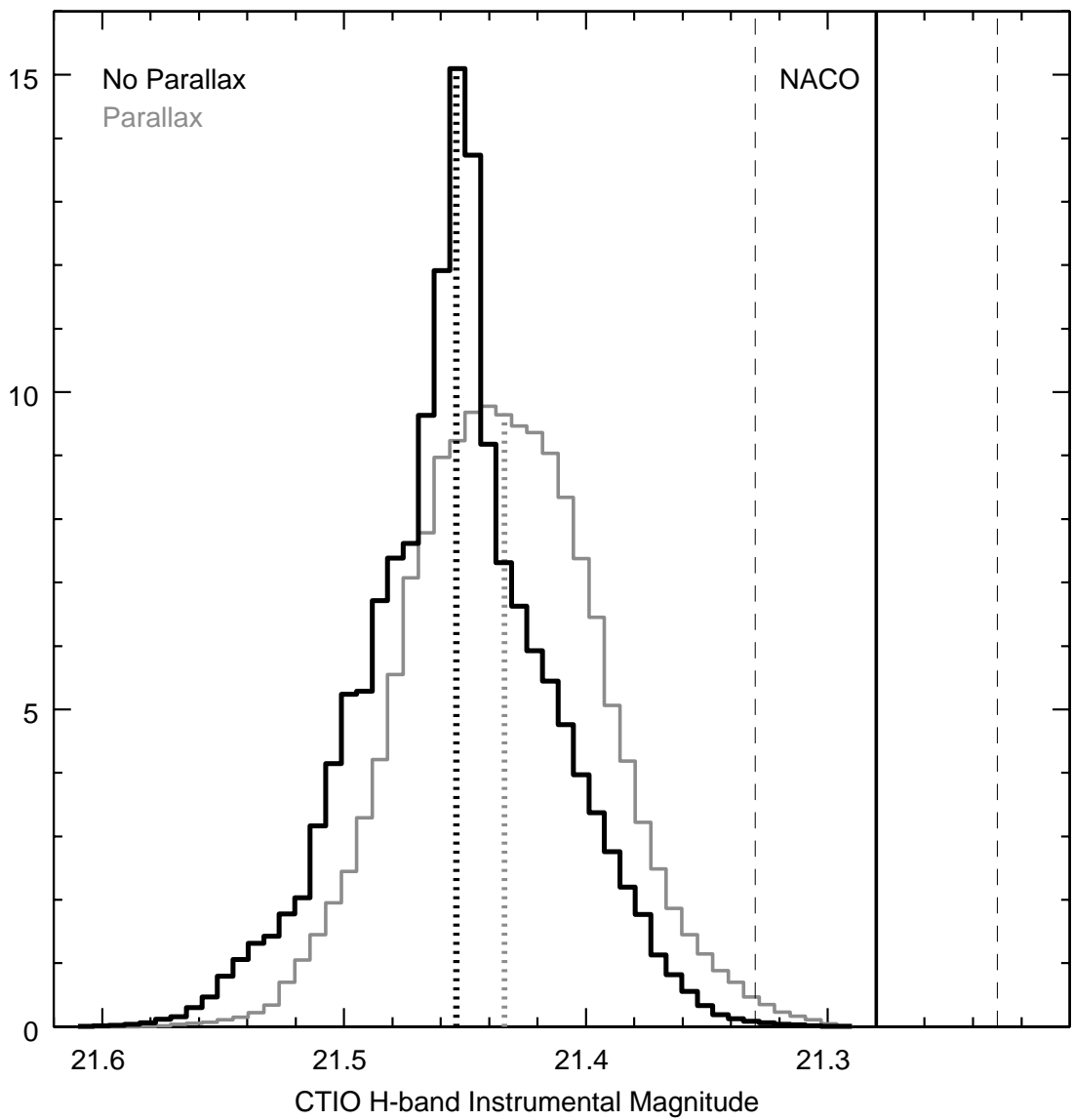}
	\caption{Probability distributions (normalized to unity) for the model-derived magnified source flux at the time of the NACO image. The curve for the case of no parallax is plotted in black and unconstrained parallax is plotted in gray. The mean for each distribution is indicated by a dotted line, and the standard deviation in each case is 0.04 mag. The best estimate of the target flux on the NACO image (21.28 mag) is marked by the vertical black line, and dashed lines are 0.05 mag conservative error bars. The error bar at 21.33 mag can be considered a robust lower limit on the amount of light detected on the NACO image.}  
	\label{fig:histograms}
\end{figure}

\begin{figure}
	\includegraphics[width=1.0 \textwidth]{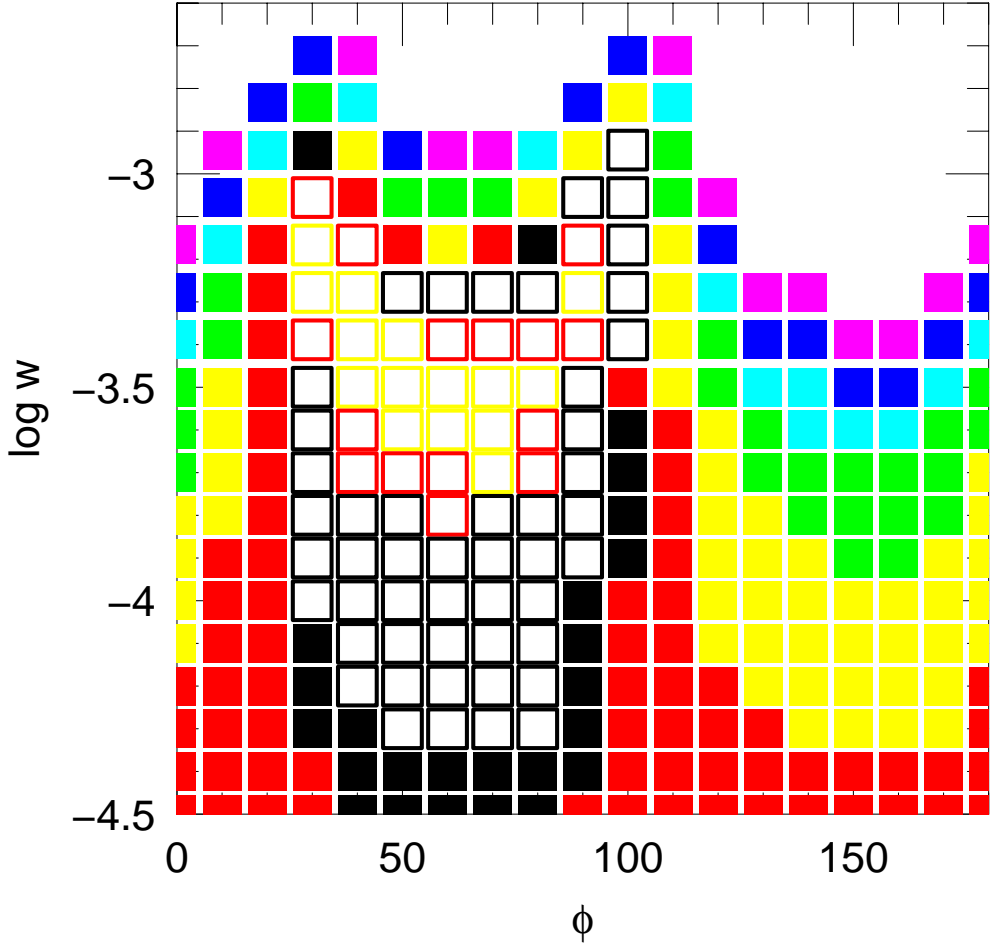}
\caption{Search for additional shear due to distant third body,
parametrized by the angle between the planet and the third body, 
$\phi$, and the width of the resulting Chang-Refsdal (1979, 1984)
caustic: $w_{\rm com} = 4 q_{\rm com} d_{\rm com}^{-2}$.  Open symbols
represent improvement relative to $\chi^2_{\rm min} -2$, i.e. the
value expected from adding two degrees of freedom to the single-planet
model.  Filled symbols indicate worse $\chi^2$.  Colors
(black, red, yellow, green, cyan, blue, magenta,white) 
= (1,4,9,16,25,49,$>49$).  There is a weak detection of shear
at $(\log w_{\rm com},\phi)=(-3.28,40^\circ)$, and a $3\,\sigma$ upper limit
$\log w_{\rm com}<-3$ over most angles.  Note that the angle between the
shear direction and the direction of source motion is 
$\phi +\alpha$, where $\alpha\simeq 69^\circ$ is the angle between
the planet direction and the direction of source motion.}
\label{fig:shear}
\end{figure}

\begin{figure}
	\includegraphics[width=1.0 \textwidth]{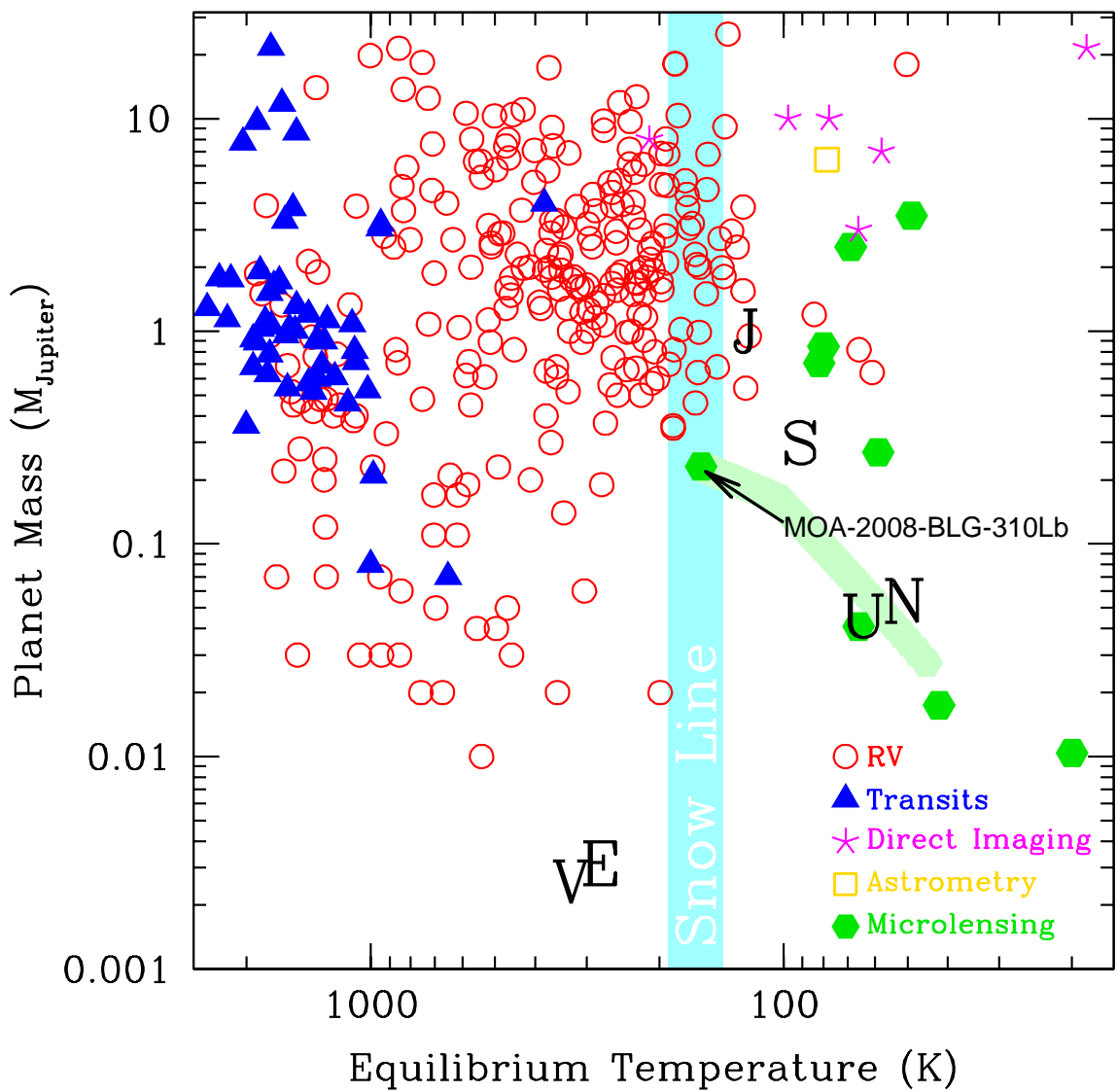}
	\caption{Mass versus equilibrium temperature for planets detected via radial velocity ({\it circles}), transits ({\it triangles}), imaging ({\it stars}), astrometry ({\it squares}), and microlensing ({\it hexagons}). If the blended light aligned with the event that was identified by VLT NACO is in fact due to the lens, MOA-2008-BLG310Lb would be the first microlensing detection to fall on the Snow Line. The tail on the marker for this detection indicates where the planet might fall if the host is a lower-mass star, rather than being identified with the blended light. (Data taken from http://exoplanet.eu/, maintained by
J.\ Schneider.)}  
	\label{fig:detections}
\end{figure}

\end{document}